\documentclass[10pt,twocolumn,letterpaper]{article}

\usepackage{cvpr}              

\usepackage{tikz}
\usetikzlibrary{positioning, shapes.multipart}
\usepackage{multirow}
\usepackage{comment}
\usepackage{float}
\usepackage{array}
\usepackage{placeins}

\newif\ifdraft
\drafttrue

\DeclareMathOperator*{\agg}{agg}
\DeclareMathOperator*{\mean}{mean}
\DeclareMathOperator*{\sumagg}{sum}

\definecolor{cvprblue}{rgb}{0.21,0.49,0.74}
\usepackage[pagebackref,breaklinks,colorlinks,allcolors=cvprblue]{hyperref}

\def\paperID{15821} 
\def\confName{CVPR}
\def\confYear{2026}

\title{PRIMU: Uncertainty Estimation for Novel Views in Gaussian Splatting from Primitive-Based Representations of Error and Coverage}

\author{Thomas Gottwald\\
Department of  Mathematics\\
University of Wuppertal\\
Wuppertal, Germany\\
{\tt\small gottwald@uni-wuppertal.de}
\and
Edgar Heinert\\
Institute of Computer Science\\
University of Osnabrück\\
Osnabrück, Germany\\
{\tt\small edgar.heinert@uni-osnabrueck.de}
\and
Peter Stehr\\
Department of  Mathematics\\
University of Wuppertal\\
Wuppertal, Germany\\
{\tt\small stehr@uni-wuppertal.de}
\and
Chamuditha Jayanga Galappaththige\\
Centre for Robotics\\
Queensland University of Technology\\
Brisbane, Australia\\
{\tt\small chamuditha.galappaththige@qut.edu.au}
\and
Matthias Rottmann\\
Institute of Computer Science\\
University of Osnabrück\\
Osnabrück, Germany\\
{\tt\small matthias.rottmann@uni-osnabrueck.de}
}

\begin{document}
\maketitle

\begin{abstract}

We introduce Primitive-based Representations of Uncertainty (PRIMU), 
a post-hoc uncertainty estimation (UE) framework for Gaussian Splatting (GS).
Reliable UE is essential for deploying GS in safety-critical domains such as robotics and medicine.
Existing approaches typically estimate Gaussian-primitive variances and rely on the rendering process to obtain pixel-wise uncertainties.
In contrast, we construct primitive-level representations of error and visibility/coverage from training views, capturing interpretable uncertainty information. 
These representations are obtained by projecting view-dependent training errors and coverage statistics onto the primitives. 
Uncertainties for novel views are inferred by rendering these primitive-level representations, producing uncertainty feature maps, which are aggregate through pixel-wise regression on holdout data. 
We analyze combinations of uncertainty feature maps and regression models to understand how their interactions affect prediction accuracy and generalization.
PRIMU also enables an effective active view selection strategy by directly leveraging these uncertainty feature maps.
Additionally, we study the effect of separating splatting into foreground and background regions.
Our estimates show strong correlations with true errors, outperforming state-of-the-art methods, especially for depth UE and foreground objects.
Finally, our regression models show generalization capabilities to unseen scenes, enabling UE without additional holdout data.

\end{abstract}    
\section{Introduction}
\label{sec:intro}

\begin{figure}
    \centering
    \begin{tabular}{ >{\centering\arraybackslash} m{2mm}  >{\centering\arraybackslash} m{3.4cm}
    @{\hspace{1.5mm}}>{\centering\arraybackslash} m{3.4cm}}
        & Color & Depth \\
        \rotatebox[origin=l]{90}{Rendering} &
        \includegraphics[width=\linewidth]{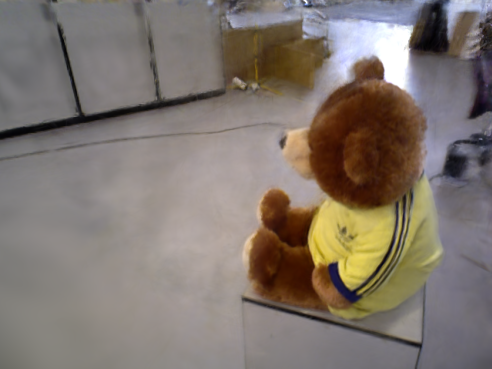} &
        \includegraphics[width=\linewidth]{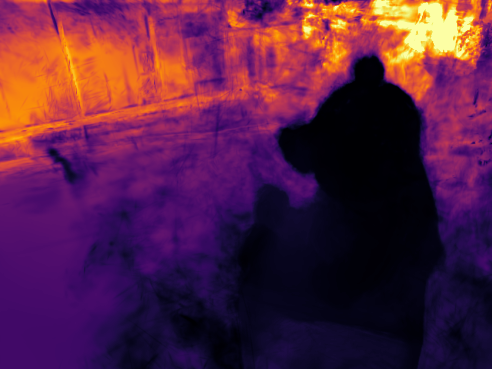} \\
        \rotatebox[origin=l]{90}{Error} &
        \includegraphics[width=\linewidth]{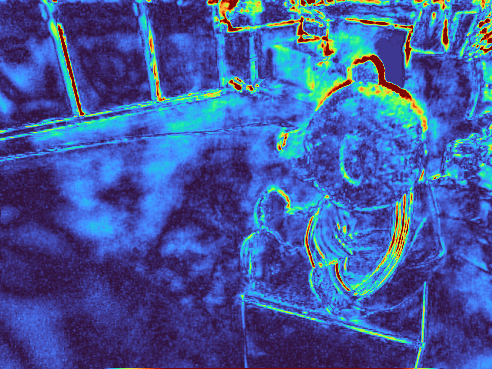} &
        \includegraphics[width=\linewidth]{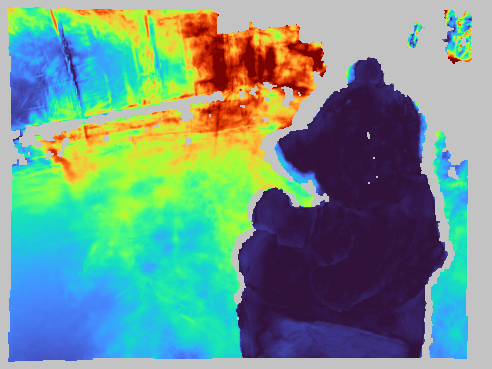} \\
        \rotatebox[origin=l]{90}{Uncertainty} &
        \includegraphics[width=\linewidth]{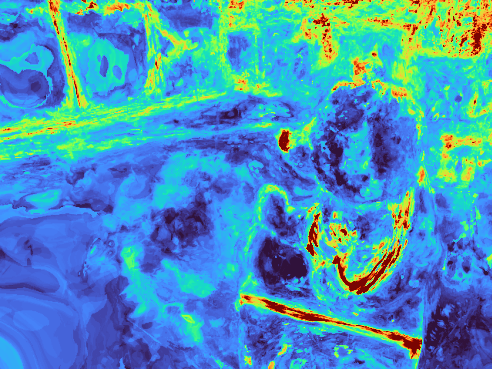} &
        \includegraphics[width=\linewidth]{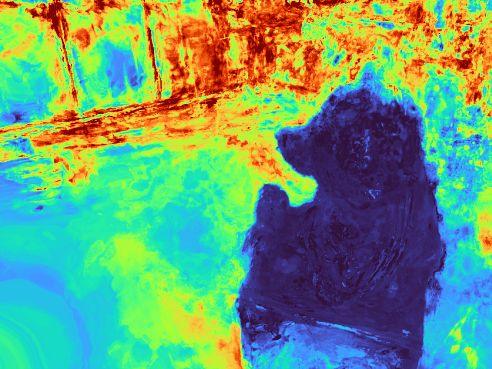}
    \end{tabular}
    \caption{
    Comparison of PRIMU uncertainty maps to error for color and depth renderings.
    Undefined depth error is represented as light gray.
    Additional examples in Appendix \ref{appen:VisualExamples}.
    }
    \label{fig:page1fig}
\end{figure}

Recent advances in volume-rendering-based techniques, which learn to represent the radiance field as a function mapping 3D positions and viewing directions to observed appearance, have driven major progress in novel view synthesis and 3D reconstruction.
Neural Radiance Fields (NeRFs)~\cite{mildenhall2021nerf} employ neural networks (NNs) for differentiable volume rendering, supervised by posed camera views~\cite{kajiya1984ray,max95opticalmodels,zwicker2001ewa}.
Gaussian Splatting (GS)~\cite{kerbl20233dgaussiansplatting} replaces NNs with 3D Gaussian primitives of learnable position, variance, and opacity.
These primitives are initialized from structure-from-motion (SfM) point clouds~\cite{snavely2006photo} and optimized via a rendering loss.
While both NeRF and GS achieve high visual fidelity~\cite{barron2022mipnerf,zhang2025evaluating}, GS offers faster rendering and a more explicit scene representation~\cite{kerbl20233dgaussiansplatting,matsuki2024gaussian}.

However, GS still faces challenges similar to NeRF-based methods.
Gaussian splats often misalign with scene geometry even when the rendered appearance looks correct~\cite{wang2021neus,remondino2023critical,huang20242dgaussiansplatting}, leading to a \emph{depth error} between the true geometry and splat distribution.
Moreover, when rendering a novel view, it is not always clear whether the resulting image accurately approximates the real scene or where potential inaccuracies occur~\cite{shen2022conditional,suenderhauf2023densityaware,aria2025modelinguncertainty}.
We refer to this mismatch as the \emph{rendering error}.
Rendering-error-based optimization can introduce geometric misalignment~\cite{jung2024relaxing}. Recent works mitigate these effects via geometry smoothing~\cite{warburg2023nerfbusters}, improved optimization~\cite{huang2024gs++}, or co-regularization of inconsistent Gaussians~\cite{zhang2024cor}.

Uncertainty estimation (UE) directly identifies high error regions in renderings.
Accurate UE enables trustworthy geometry recovery, which is critical in robotics and medical imaging~\cite{thrun2002probabilistic,long2021learning,shamsi2021uncertainty}.
Selecting new training views with high uncertainty can improve reconstruction quality while reducing data requirements~\cite{shen2022conditional,Jiang2023FisherRF}, and can guide robotic path planning~\cite{Jiang2023FisherRF}.
Yet, UE in GS remains an open challenge.
Existing methods usually predict per-primitive uncertainty and render pixel-level uncertainty maps~\cite{Jiang2023FisherRF,wilson2024modelinguncertainty3dgaussian}, using either Fisher information~\cite{Jiang2023FisherRF} or learned distributions over primitives~\cite{aria2025modelinguncertainty,wilson2024modelinguncertainty3dgaussian,li2024variational}.
Whether these estimates reflect depth or rendering errors depends on the underlying formulation, scene geometry~\cite{Jiang2023FisherRF}, and incorporated semantics~\cite{wilson2024modelinguncertainty3dgaussian}.

We propose PRIMU (Primitive-based Representations of Uncertainty), a novel framework for post-hoc UE in GS.
PRIMU represents uncertainty using Gaussian primitive representations that capture training-view coverage and reconstruction error.
We model these representations with and without directional dependency, enabling camera-view-aware UE. 
Each rendered representation produces an \emph{uncertainty feature map} for a given pose, from which we derive pixel-wise uncertainty estimates using gradient boosting regression~\cite{natekin2013gradient} trained on one or a few novel views.
Examplary pixel-wise uncertainty maps are presented in \cref{fig:page1fig}.

Our experiments evaluate various combinations of uncertainty feature maps and regression models, showing how their interactions affect prediction accuracy and generalization.
We also demonstrate that the uncertainty feature maps themselves can directly guide active view selection (AVS), improving reconstruction quality and data efficiency without the need of holdout data and training an explicit UE regressor.
The regression models generalize well to unseen views achieving high correlation with true errors, and also demonstrate generalization capabilities to new scenes.
This indicates that our uncertainty feature map construction captures the UE relevant information on a low dimensionality.
Separating scenes into foreground and background reveals especially high UE performance for foreground objects.

We summarize our main contributions as follows:
\begin{itemize}
    \item 
    We propose a post-hoc UE method for GS based on Gaussian primitive representations of training-view error and coverage that achieves state-of-the-art performance for RGB and depth UE.
    \item 
    We perform an extensive ablation study exploring different combinations of uncertainty feature maps and regression models, revealing how feature interactions influence prediction accuracy and generalization across scenes.
    \item 
    We use the proposed uncertainty feature maps directly to guide AVS, improving reconstruction quality and data efficiency by prioritizing viewpoints with high uncertainty indicators.
\end{itemize}

\section{Related Work}
\label{sec:relatedwork}

\begin{figure*}
    \centering
    \scalebox{0.82}{\input{images/03_method/flow_chart}}
    \caption{
    PRIMU projects rendering errors, observation coverage, and field-of-view statistics from training views onto Gaussian primitives.
    From these, it creates Gaussian primitive representations that can be rendered from novel viewpoints to produce uncertainty feature maps. 
    Pixel-wise regression then predicts per-pixel errors as uncertainty estimates.
    Visual examples are shown in \cref{fig:page1fig} and Appendix \ref{appen:VisualExamples}.
    }
    \label{fig:method_outline}
\end{figure*}

We review UE approaches for NeRFs and GS, as well as their use in active view selection for novel view synthesis.

\paragraph{Uncertainty Estimation for NeRFs}

In NeRF-based methods, uncertainty is typically defined per pixel, with no intermediate representation comparable to Gaussian primitives.
Several works model stochastic radiance fields, estimating pixel-wise variance from sampled renderings~\cite{shen2022conditional,seo2023flipnerf}, while Suenderhauf et~al.~\cite{suenderhauf2023densityaware} employ NN ensembles.
Goli et~al.~\cite{goli2024bayesrays} propose a Bayesian post-processing scheme where uncertainty measures how much the radiance field can vary without degrading reconstruction, highlighting underconstrained regions.
Nakayama et~al.~\cite{nakayama2024provnerf} introduce a stochastic provenance field capturing viewing directions and estimate geometric uncertainty via triangulation~\cite[Ch.~12.6]{hartley2004multiple}.
Transferring these methods to GS is not straightforward due to its different scene representation.
Goli et~al.’s Bayesian approach has been adapted in spirit for GS by Jiang et al.~\cite{Jiang2023FisherRF}, while Nakayama et~al.~show a proof-of-concept GS experiment, though details and code are unavailable.
Additional UE research focuses on calibration~\cite{amini2025instant} and training-data uncertainty~\cite{martin2021nerfinthewild}, the latter also extended to GS~\cite{kulhanek2024wildgaussians}.
These works improve reliability or data handling, whereas our focus is on estimating uncertainties of the GS radiance-field approximation itself.

\paragraph{Uncertainty Estimation for Gaussian Splatting}

Most UE methods for GS estimate uncertainty per Gaussian primitive.
Wilson et~al.~\cite{wilson2024modelinguncertainty3dgaussian} use Bayesian post-processing to learn semantic distributions over primitives, with their variance as semantic uncertainty.
Other methods~\cite{aria2025modelinguncertainty,li2024variational,lyu2024manifold} model stochastic radiance fields and derive pixel-level uncertainties from multi-sample variance.
Li et~al.~learn fine-scale stochastic offsets for key primitives, and Lyu et~al.~reduce sampling cost via a low-dimensional manifold.
These mainly address the runtime overhead of stochastic rendering.
A pre-print has also integrated direction-dependent uncertainty directly into GS training~\cite{han2025viewdependentuncertainty}.
Jiang et~al.~\cite{Jiang2023FisherRF} estimate parameter variances using Fisher information, conceptually similar to Goli et~al.~\cite{goli2024bayesrays}.
Their and Wilson et~al.’s~\cite{wilson2024modelinguncertainty3dgaussian} methods render pixel-level uncertainties by replacing color with Gaussian-level uncertainty values.
Our method differs by operating entirely as post-processing, requiring no GS retraining.
In our experiments, we demonstrate that training a lightweight regression model with a few views enables generalization to novel views and, to some extent, unseen scenes.
Unlike~\cite{Jiang2023FisherRF}, which uses a single uncertainty measure, PRIMU extracts 13 optionally direction-dependent Gaussian-primitive features from training-view coverage and reconstruction error.
Rendering these as uncertainty feature maps and regressing them per pixel with gradient boosting yields accurate, generalizable depth and rendering uncertainty.

\paragraph{Active View Selection}

AVS uses UE to choose additional views that improve reconstruction while minimizing training data.
For UE methods, new views are typically selected from a set of candidate viewpoints by maximizing uncertainty.
This has been demonstrated for both NeRF~\cite{goli2024bayesrays,suenderhauf2023densityaware} and GS~\cite{li2024variational,lyu2024manifold} UE approaches.
Jiang et~al.~\cite{Jiang2023FisherRF} quantify the information gain of new views using the same principle underlying their GS UE method of Fisher information.
A similar idea is applied by Pan et~al.~\cite{pan2022activenerf} for NeRF, who model information gain as a variance reduction relative to a prior Gaussian color distribution.
Kopanas and Drettakis~\cite{kopanas2023improvingnerf} instead select new views based on observation frequency and angular uniformity.
In contrast to~\cite{goli2024bayesrays,suenderhauf2023densityaware,li2024variational,lyu2024manifold}, we do not use our uncertainty estimates directly, but find a training view coverage map effective for AVS.
This allows to efficiently prioritize high-uncertainty viewpoints, without the need to train a regression model.
Unlike~\cite{Jiang2023FisherRF,pan2022activenerf}, we do not explicitly estimate information gain, yet our approach surpasses the GS AVS method of~\cite{Jiang2023FisherRF} in their own evaluation setting.

\section{Method}
\label{sec:method}

In this section, we first review GS volume rendering before introducing our Gaussian primitive representations for UE. We then describe how these representations are constructed and rendered to obtain uncertainty feature maps, which are used for UE in novel views.
In \cref{fig:method_outline} we show the outline of the PRIMU method for UE.

\paragraph{Gaussian Splatting}
GS approximates a radiance field using a set of Gaussian primitives.  
Each primitive is modeled as a scaled 3D Gaussian distribution with mean \(\mu\), covariance matrix \(\Sigma\), scalar opacity factor \(g\), and color \(c\).  
To render a view, defined by its camera position, viewing direction \(\vec{d}\), and intrinsics, all Gaussian primitives are first sorted from front to back and projected onto the image plane.
This projection, known as \emph{splatting}~\cite{zwicker2001ewa}, converts the 3D Gaussian distributions into 2D Gaussians on the image plane.

The opacity of a Gaussian primitive at pixel \(\bar{x}\) in the rendered view is given by
\begin{equation}
    \alpha(\bar{x}) = g \, \Phi(\bar{x}; \bar{\mu}, \bar{\Sigma}),
    \label{eq:GSopacity}
\end{equation}
where \(\Phi\) denotes the 2D Gaussian density with mean \(\bar{\mu}\) and covariance \(\bar{\Sigma}\).
The final color at each pixel is computed by aggregating contributions from all projected Gaussians in front-to-back order:
\begin{equation}
    C(\bar{x}) = \sum_{k=1}^K c_k(\vec{d}) \, \alpha_k(\bar{x}) \, T_k(\bar{x}),
    \label{eq:GSrendering}
\end{equation}
where \(c_k(\vec{d})\) and \(\alpha_k(\bar{x})\) denote the color and opacity of the \(k\)th Gaussian, and
\(T_k(\bar{x}) = \prod_{j=1}^{k-1} (1 - \alpha_j(\bar{x}))\)
represents the accumulated transparency from primitives \(1\) to \((k-1)\).

The color \(c_k(\vec{d})\) depends on the view direction \(\vec{d}\), which is encoded using spherical harmonics~\cite{muller2006spherical,price:s2fft}.
GS learns all Gaussian parameters (mean, covariance, opacity, and color)
through gradient descent on a reconstruction loss computed over a set of posed training views.

\paragraph{Gaussian Primitive Representations}
Our Gaussian primitive representations capture uncertainty from two complementary sources: (1) limited training-view coverage and (2) rendering errors in the training views.  
These representations are computed in a post-processing step after GS training, without modifying the rendering or optimization.

We begin with a simple field-of-view (FoV) count representation that records how many training views observe each Gaussian primitive:
\begin{equation}
    F(k) = |\Omega_k|,
    \label{eq:fovCounter}
\end{equation}
where \(\Omega_k\) denotes the set of training views in which the \(k\)th Gaussian primitive lies within the viewing cone.

To represent how well a Gaussian is covered in the training data, we use its contribution factors \(\alpha_k(\bar{x}) T_k(\bar{x})\) (from Eq.~\ref{eq:GSrendering}) over the visible pixels in each training view.
These are aggregated using \(\agg \in \{\max, \sumagg, \mean\}\).
We then take the maximum aggregated value across all training views to form the training-view coverage representation:
\begin{equation}
    V(k) = \max_{v \in \Omega} \agg_{\bar{x} \in v(k)} \alpha_k(\bar{x}) T_k(\bar{x}),
    \label{eq:visRep}
\end{equation}
where \(v(k)\) denotes the pixels of view \(v\) in which the opacity of the \(k\)th Gaussian exceeds a small threshold.
This captures how strongly each Gaussian primitive is covered in at least one training view.

Analogously, we define the rendering-error representation by weighting each contributing term by the pixel-wise reconstruction error \(e_{\bar{x}}\):
\begin{equation}
    E(k) = \mean_{v \in \Omega} \agg_{\bar{x} \in v(k)} e_{\bar{x}} \, \alpha_k(\bar{x}) T_k(\bar{x}),
    \label{eq:errRep}
\end{equation}
where \(e_{\bar{x}}\) is the per-pixel \(\ell_1\) reconstruction error.
This representation captures how much a Gaussian contributes to the overall rendering error.  
Both coverage and error representations can omit the opacity term \(\alpha_k(\bar{x})\) to highlight low-opacity Gaussians that still influence reconstruction quality.

\paragraph{Direction-Dependent Representations}
We extend both representations to be direction-dependent by encoding how the coverage or error of each Gaussian varies with viewing direction.
To this end, we employ the von Mises-Fisher distribution~\cite{fisher1993statistical,mardia2009directional}, a spherical analogue of the normal distribution, defined as
\begin{equation}
    f(\vec{x}; \vec{\nu}, \kappa) = \frac{\kappa}{2\pi \sinh(\kappa)} e^{\kappa \, \vec{\nu}^\top \vec{x}},
\end{equation}
where \(\vec{\nu}\) is the mean direction and \(\kappa\) controls the concentration around it.
We rescale this density so its maximum equals one and use it to weight contributions in Eqs.~\ref{eq:visRep} and~\ref{eq:errRep}.
The resulting direction-dependent representations are
\begin{align}
    V^*(k, \vec{d}) &= \max_{v \in \Omega} \agg_{\bar{x} \in v(k)} 
        \alpha_k(\bar{x}) T_k(\bar{x}) 
        e^{\kappa \vec{\nu}_v^\top \vec{d} - \kappa},
    \label{eq:visRepDirDep} \\
    E^*(k, \vec{d}) &= \mean_{v \in \Omega} \agg_{\bar{x} \in v(k)}
        e_{\bar{x}} \, \alpha_k(\bar{x}) T_k(\bar{x}) 
        e^{\kappa \vec{\nu}_v^\top \vec{d} - \kappa},
    \label{eq:errRepDirDep}
\end{align}
where \(\vec{\nu}_v\) denotes the direction of training view \(v\) and \(\vec{d}\) the direction at which the representation is evaluated.
Intuitively, this corresponds to placing a scaled von Mises-Fisher distribution on the sphere around each Gaussian primitive at the directions of its observed training views.
For efficient rendering, these direction-dependent representations are encoded using spherical harmonics, consistent with the color encoding in GS.

\paragraph{Regression Models}
To obtain pixel-wise uncertainty estimates for novel views, we render a set of chosen \emph{uncertainty feature maps} by replacing the Gaussian color term in Eq.~\ref{eq:GSrendering} with the respective Gaussian primitive representation.
The resulting feature maps serve as input to a pixel-wise regression model that predicts rendering or depth error.

The regression model is trained on a small set of hold-out views, using the true rendering or depth error as the target
\begin{equation}
    e_{\bar{x}} = \| R(\bar{x}) - \hat{R}(\bar{x}) \|_1.
\end{equation}
\(R(\bar{x})\) and \(\hat{R}(\bar{x})\) denote the ground-truth and rendered values for pixel \(\bar{x}\), respectively.
Once trained, this model can be applied to novel viewpoints or scenes to produce dense, pixel-wise uncertainty maps for both color and depth.

\section{Numerical Results}
\label{sec:results}

\begin{figure*}
    \centering
    \footnotesize
    \setlength{\tabcolsep}{0pt}
    \begin{tabular}{>{\centering\arraybackslash} m{0.142\linewidth} >{\centering\arraybackslash} m{0.142\linewidth} >{\centering\arraybackslash} m{0.142\linewidth} >{\centering\arraybackslash} m{0.142\linewidth} >{\centering\arraybackslash} m{0.142\linewidth} >{\centering\arraybackslash} m{0.142\linewidth} >{\centering\arraybackslash} m{0.142\linewidth}}
         Ground Truth & Rendering & \(\ell_1\) Error Map & PRIMU* & manifold & var3DGS & FisherRF
    \end{tabular}
    \includegraphics[width=\linewidth,trim={0 0.5cm 0 0.5cm},clip]{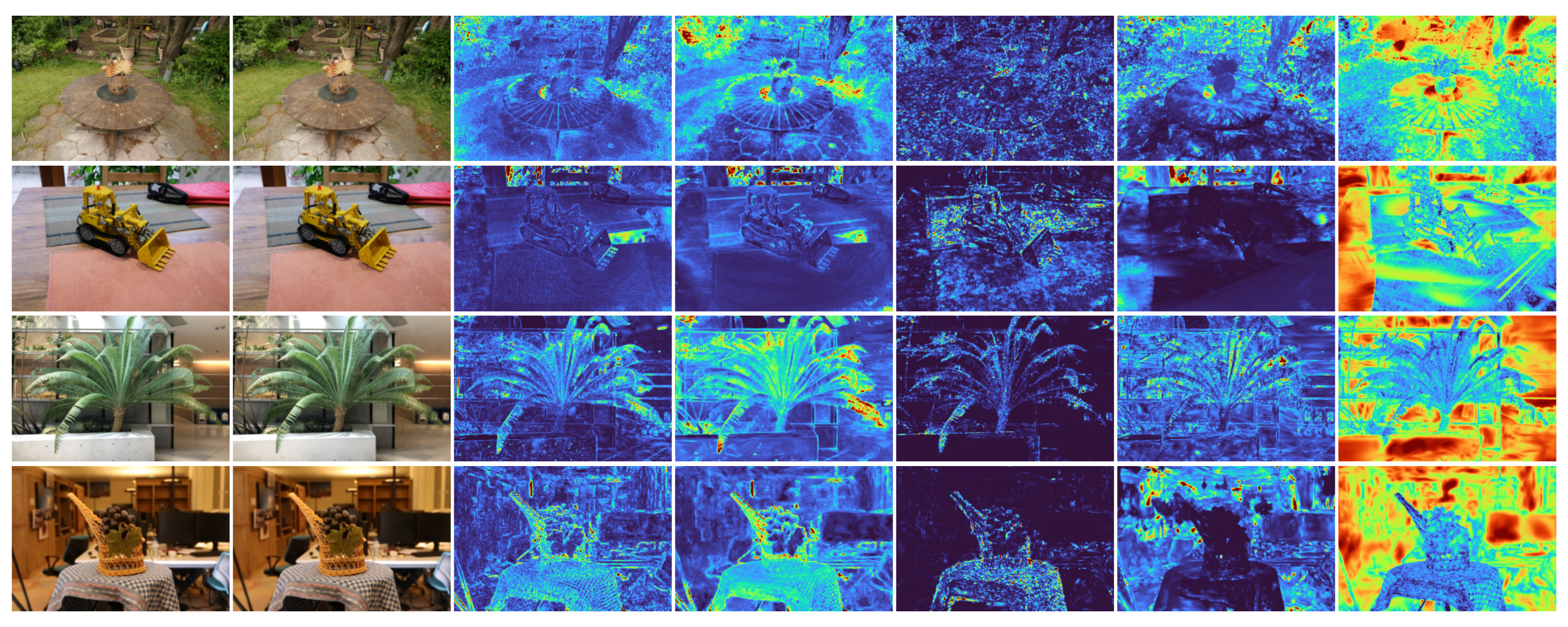}
    \caption{Qualitative comparison of uncertainty feature maps of FisherRF~\cite{Jiang2023FisherRF}, manifold~\cite{lyu2024manifold}, var3DGS~\cite{li2024variational} and direction-dependent PRIMU* (ours), to the \(\ell_1\) rendering error.
    Scenes from top to bottom: MipNeRF360 garden and kitchen, LLFF fern and LF basket.}
    \label{fig:qualtiative_comparision}
\end{figure*}

\begin{table*}
    \setlength{\tabcolsep}{4pt}
    \centering
    \scalebox{0.85}{
    \begin{tabular}{l|ll|lll|ll|lll|ll|lll}
\toprule
 & \multicolumn{5}{c}{LF} & \multicolumn{5}{|c}{MipNeRF360} & \multicolumn{5}{|c}{LLFF} \\
 & AUSE & Pears. & PSNR & SSIM & LPIPS & AUSE & Pears. & PSNR & SSIM & LPIPS & AUSE & Pears. & PSNR & SSIM & LPIPS \\
method & (\(\downarrow\)) & (\(\uparrow\)) & (\(\uparrow\)) & (\(\uparrow\)) & (\(\downarrow\)) & (\(\downarrow\)) &(\(\uparrow\)) & (\(\uparrow\)) & (\(\uparrow\)) & (\(\downarrow\)) & (\(\downarrow\)) & (\(\uparrow\)) & (\(\uparrow\)) & (\(\uparrow\)) & (\(\downarrow\)) \\
\midrule
FisherRF & 0.753 & -0.13 & 34.18 & 0.969 & 0.043 & 0.738 & -0.054 & 27.508 & 0.817 & 0.215 & 0.988 & -0.088 & 23.814 & 0.801 & 0.226 \\
maniflod & 0.435 & 0.099 & 33.712 & 0.969 & 0.044 & 0.52 & 0.07 & 27.273 & 0.809 & 0.23 & 0.545 & 0.071 & 24.239 & 0.801 & 0.223 \\
var3DGS & 0.688 & 0.102 & 23.101 & 0.826 & 0.233 & 0.52 & 0.173 & 26.039 & 0.762 & 0.298 & 0.476 & 0.293 & 25.778 & 0.832 & 0.204 \\
PRIMU (ours) & 0.291 & 0.337 & 34.18 & 0.969 & 0.043 & 0.391 & 0.203 & 27.508 & 0.817 & 0.215 & 0.284 & 0.363 & 23.814 & 0.801 & 0.226 \\
PRIMU*(ours) & \textbf{0.286} & \textbf{0.351} & 34.18 & 0.969 & 0.043 & \textbf{0.378} & \textbf{0.236} & 27.508 & 0.817 & 0.215 & \textbf{0.281} & \textbf{0.368} & 23.814 & 0.801 & 0.226 \\
\bottomrule
\end{tabular}

    }
    \caption{
    Rendering error UE results (AUSE, Pears.) on LF, MipNeRF360, and LLFF. PRIMU* uses direction-dependent coverage and error maps. Post-hoc UE methods don't affect GS splatting quality (PSNR, SSIM, LPIPS).
    }
    \label{tab:rgbResults}
\end{table*}

\begin{table}
    \centering
    \scalebox{0.85}{
    \begin{tabular}{l|ll|lll}
\toprule
 & \multicolumn{5}{c}{LF} \\
 & AUSE & Pears. & PSNR & SSIM & LPIPS \\
method & (\(\downarrow\)) & (\(\uparrow\)) & (\(\uparrow\)) & (\(\uparrow\)) & (\(\downarrow\)) \\
\midrule
BayesRays & 0.173 & 0.243 & 25.181 & 0.887 & 0.077 \\
FisherRF & 0.305 & 0.105 & 34.18 & 0.969 & 0.043 \\
maniflod & 0.436 & 0.036 & 33.712 & 0.969 & 0.044 \\
PRIMU (ours) & \textbf{0.115} & 0.722 & 34.18 & 0.969 & 0.043 \\
PRIMU*(ours) & 0.118 & \textbf{0.728} & 34.18 & 0.969 & 0.043 \\
\bottomrule
\end{tabular}
    }
    \caption{
    Depth UE results (AUSE, Pearson) on LF. PRIMU* uses direction-dependent coverage and error maps. Post-hoc UE methods don't affect GS splatting quality (PSNR, SSIM, LPIPS).
    }
    \label{tab:dephtResults}
\end{table}

\begin{table*}
    \setlength{\tabcolsep}{2pt}
    \centering
    \scalebox{0.85}{
    \begin{tabular}{l|ll|ll|ll|ll|ll|ll|ll|ll|ll}
\toprule
& \multicolumn{2}{r|}{predicted error type} & \multicolumn{8}{c|}{rendering error} & \multicolumn{8}{c}{depth error} \\
& \multicolumn{2}{r|}{evaluation on} & \multicolumn{4}{c|}{entire views} & \multicolumn{4}{c|}{object-centric} & \multicolumn{4}{c|}{entire views} & \multicolumn{4}{c}{object-centric} \\
 &  & dataset & \multicolumn{2}{c|}{LF} & \multicolumn{2}{c|}{TUM} & \multicolumn{2}{c|}{LF} & \multicolumn{2}{c|}{TUM} & \multicolumn{2}{c|}{LF} & \multicolumn{2}{c|}{TUM} & \multicolumn{2}{c|}{LF} & \multicolumn{2}{c}{TUM} \\
 &  & metric & AUSE & Pears. & AUSE & Pears.  & AUSE & Pears.  & AUSE & Pears.  & AUSE & Pears.  & AUSE & Pears.  & AUSE & Pears.  & AUSE & Pears.  \\
\multirow[t]{2}{*}{\rotatebox{90}{views}} & method & model & (\(\downarrow\)) & (\(\uparrow\)) & (\(\downarrow\)) & (\(\uparrow\)) & (\(\downarrow\)) & (\(\uparrow\)) & (\(\downarrow\)) & (\(\uparrow\)) & (\(\downarrow\)) & (\(\uparrow\)) & (\(\downarrow\)) & (\(\uparrow\)) & (\(\downarrow\)) & (\(\uparrow\)) & (\(\downarrow\)) & (\(\uparrow\)) \\
\midrule
 & FisherRF &  & 0.878 & -0.224 & 0.957 & -0.241 & 0.575 & 0.129 & 0.249 & 0.458 & 0.322 & 0.31 & 0.742 & -0.08 & 0.63 & 0.05 & 0.38 & 0.303 \\
\cline{1-19} \cline{2-19}
\multirow[t]{6}{*}{\rotatebox[origin=r]{90}{1 view}} & \multirow[t]{2}{*}{reg. FisherRF} & grad. & 0.282 & 0.318 & 0.435 & 0.181 & 0.119 & 0.704 & 0.169 & 0.645 & 0.133 & 0.683 & 0.328 & 0.375 & 0.135 & 0.698 & 0.228 & 0.521 \\
 &  & lin. & 0.332 & 0.182 & 0.404 & 0.143 & 0.421 & 0.404 & 0.292 & 0.523 & 0.212 & 0.452 & 0.363 & 0.305 & 0.398 & 0.398 & 0.453 & 0.364 \\
\cline{2-19}
 & \multirow[t]{2}{*}{PRIMU (ours)} & grad. & 0.244 & 0.412 & 0.416 & 0.217 & 0.072 & \underline{0.822} & 0.101 & 0.756 & 0.112 & 0.734 & 0.213 & 0.536 & 0.071 & \underline{0.835} & 0.1 & 0.745 \\
 &  & lin. & 0.286 & 0.309 & 0.353 & 0.29 & 0.254 & 0.633 & 0.248 & 0.617 & 0.202 & 0.545 & 0.188 & 0.571 & 0.246 & 0.707 & 0.282 & 0.586 \\
\cline{2-19}
 & \multirow[t]{2}{*}{PRIMU*(ours)} & grad. & 0.238 & 0.419 & 0.415 & 0.23 & 0.071 & 0.818 & 0.114 & 0.724 & 0.108 & 0.739 & 0.239 & 0.501 & \underline{0.07} & 0.831 & 0.11 & 0.718 \\
 &  & lin. & 0.276 & 0.35 & 0.362 & 0.296 & 0.376 & 0.526 & 0.327 & 0.526 & 0.202 & 0.557 & 0.226 & 0.521 & 0.313 & 0.612 & 0.427 & 0.435 \\
\cline{1-19} \cline{2-19}
\multirow[t]{3}{*}{\rotatebox[origin=r]{90}{m-view}} & reg. FisherRF & grad. & 0.26 & 0.352 & 0.36 & 0.267 & 0.098 & 0.751 & 0.102 & 0.766 & 0.127 & 0.715 & 0.226 & 0.515 & 0.1 & 0.757 & 0.121 & 0.7 \\
\cline{2-19}
 & PRIMU (ours) & grad. & \underline{0.222} & \underline{0.446} & \underline{0.335} & \underline{0.302} & \underline{0.06} & \textbf{0.842} & \textbf{0.06} & \textbf{0.851} & \underline{0.099} & \underline{0.777} & \textbf{0.132} & \textbf{0.653} & \textbf{0.052} & \textbf{0.859} & \textbf{0.062} & \textbf{0.849} \\
\cline{2-19}
 & PRIMU*(ours) & grad. & \textbf{0.218} & \textbf{0.458} & \textbf{0.329} & \textbf{0.321} & \textbf{0.059} & \textbf{0.842} & \underline{0.069} & \underline{0.811} & \textbf{0.097} & \textbf{0.788} & \underline{0.161} & \underline{0.605} & \textbf{0.052} & \textbf{0.859} & \underline{0.068} & \underline{0.825} \\
\cline{1-19} \cline{2-19}
\bottomrule
\end{tabular}

    }
    \caption{
    Scene separation UE results. “entire views” evaluates on the full GS scene; “object-centric” limits to the central object. PRIMU* uses direction-dependent coverage and error maps.
    }
    \label{tab:obejctStudyResults}
\end{table*}

This section presents our experiments on UE and AVS.
First, we describe the datasets used, followed by the implementation details for the uncertainty feature maps.
Next, we discuss the setup of our UE experiments, detailing the base configuration used for our method. We also cover the UE metrics and baseline methods used for comparison.
We then present the numerical results of our UE experiments, an UE scene separation study, and a study on combining uncertainty feature maps.
Finally, we conclude this section with our AVS setup and experiments.

\paragraph{Datasets}

We evaluate our approach on four datasets:
MipNeRF360~\cite{barron2022mipnerf}, Local Light Field Fusion (LLFF)~\cite{mildenhall2019locallightfieldfusion}, Light Field (LF)~\cite{yuecer2016efficient3d,shen2022conditional}, and eleven object-centric scenes from the TUM RGB-D SLAM dataset (TUM)~\cite{sturm12iros}.
The MipNeRF360 dataset contains nine 360° scenes (four indoor and five outdoor), while the LLFF dataset comprises eight front-facing scenes. From LF, we use four scenes (three indoor and one outdoor) with available ground-truth depth. From TUM, we use object-centric scenes with ground-truth depth for all views.
This ground-truth depth may not be defined for all pixels.
The TUM images often include motion blur, i.e.\ inconsistent 2D representations of 3D scenes, which lead to higher rendering errors.
We use MipNeRF360, LLFF, and LF for rendering UE and LF for depth UE.
For the scene separation study, performed for rendering and depth UE, we use LF and TUM. AVS experiments are conducted on the MipNeRF360 dataset.

\paragraph{Uncertainty Feature Maps}
We employ a) \(13\) direction-independent uncertainty feature maps: one FoV counter map, six coverage maps, and six error maps derived from the representations in \cref{eq:fovCounter,eq:visRep,eq:errRep}.  
The coverage and error maps are computed using aggregation functions \(\agg\in\{\max, \sumagg, \mean\}\), both including (\(V_{\agg}\), \(E_{\agg}\)) and excluding (\(V_{\agg}^{(\alpha=1)}\), \(E_{\agg}^{(\alpha=1)}\)) opacity \(\alpha_k(\bar{x})\). Gaussian primitives are considered visible when \(\alpha_k(\bar{x}) \geq 1/255\) and \(T_k(\bar{x}) \geq 10^{-3}\).  
For comparison we use b) the same set of feature maps with direction-dependent Gaussian primitive features (\cref{eq:visRepDirDep,eq:errRepDirDep}) using a von Mises-Fisher distribution with \(\kappa=8\) and spherical harmonics of degree four.

\paragraph{Experimental Setup for Uncertainty Estimation}
Unless stated otherwise, pixel-wise UE is performed via gradient boosting regression~\cite{natekin2013gradient} trained on one hold-out view per scene to predict rendering or depth error.
Separate regressors are trained for rendering and depth.
UE evaluations are always performed on views that are not used for training the splatting or meta-regression models.
In a scene separation study, we additionally train regressors on object and background regions using Gaussian Grouping~\cite{Ye2024grouping}.

UE performance is evaluated using the Area Under the Sparsification Error (AUSE)~\cite{ilg2018uncertainty} and Pearson correlation of the uncertainty maps to the error maps (Pears.).
AUSE measures the area between filtration curves for mean absolute error (MAE) as high-error or high-uncertainty pixels are removed. Lower values indicate better alignment between uncertainty and error. We normalize total error per view to one for comparability. Pearson correlation complements AUSE with a direct measure of linear dependency, as standard regression metrics like coefficient of determination (\(R^2\)) are not applicable due to shifts in input and target ranges.
We also report the reconstruction quality metrics Peak Signal-to-Noise Ratio (PSNR), Structural Similarity Index Measure (SSIM)~\cite{wang2004imagequality}, and Leaned Perceptual Image Patch Similarity (LPIPS)~\cite{zhang2018unreasonable} to provide context on the underlying radiance field quality.

We compare our method against the NeRF UE method by Gaoli et al.~\cite{goli2024bayesrays} (BayesRays) and the GS UE methods by Jiang et al.~\cite{Jiang2023FisherRF} (FisherRF), Li et al.~\cite{li2024variational} (var3DGS) and Lyu et al.\cite{lyu2024manifold} (manifold).
BayesRays provides post-hoc UE for NeRFs and applies only to depth UE.  
Var3DGS and manifold introduce stochasticity in GS by learning distributions over Gaussian parameters and derive uncertainties from multiple sampled renderings.
FisherRF computes Gaussian primitive uncertainty via the diagonal of the Fisher information matrix and is the only post-hoc GS UE baseline.
In the scene separation study the post-hoc UE method FisherRF is the only applicable method to the splatting containing semantic information produced by Gaussian Grouping~\cite{Ye2024grouping}.
Therefore, we introduce a regression-based variant of FisherRF for this study that uses the same regression setup as our method, but with six Fisher-derived uncertainty maps as inputs (details in Appendix~\ref{appen:fisherRFbaseline}).

\paragraph{Uncertainty Estimation Results}
Regression models are trained on a single hold-out view and evaluated on the remaining views. Separate regression models are used for direction-independent and direction-dependent feature maps.
As shown in \cref{tab:rgbResults}, our method clearly outperforms FisherRF, manifold and var3DGS for rendering UE, and achieves the best AUSE and correlation scores while being purely post-processing and therefore not degrading rendering quality.
In terms of rendering quality we mainly see similar qualities for all methods except for var3DGS on LF, where the quality is noticeably lower.
Interestingly, we observe for the forward facing LLFF dataset that the stochastic GS methods used by var3DGS and manifold have slightly higher rendering quality than standard GS~\cite{kerbl20233dgaussiansplatting}, which is used to produce the splattings used by FisherRF and PRIMU.
Direction-dependent features provide a small but consistent benefit for rendering UE, likely due to view-dependent color effects captured by GS.
\Cref{fig:qualtiative_comparision} shows a qualitative comparison of the uncertainty feature maps and \(\ell_1\) error for our method, using direction-dependent feature maps, as well as for the baseline methods.
In depth UE (\cref{tab:dephtResults}), our approach again performs best, surpassing BayesRays and FisherRF.
BayesRays as a NeRF based method has significantly lower rendering quality compared to the 3DGS methods, which undercuts its good AUSE scores.
For the depth UE setting, direction-dependence has little effect, suggesting that depth uncertainty is more direction-invariant in itself.

\paragraph{Uncertainty Estimation Scene Separation Study}
We analyze UE performance when both trained and evaluated on only the central object versus the entire scene (see \cref{tab:obejctStudyResults}).
In Appendix~\ref{appen:UEsceneSeparation} we present the results for UE on scene backgrounds.
Regression models are trained on one or several hold-out views, using both gradient boosting and linear regression.  
The regression-based FisherRF baseline performs consistently better than the original, confirming the benefit of meta-regression.
Our method further improves upon regression-based FisherRF across all settings, demonstrating the superior utility of our uncertainty feature maps.
Gradient boosting performs better than linear regression for both our approach and regression-based FisherRF.
This is particularly evident in object-centric setups, where gradient boosting significantly outperforms linear regression, suggesting stronger nonlinear relationships in this regime.
Compared to the entire-scene setting, we observe significantly improved UE in the object-centric setting for both our PRIMU and the regression-based FisherRF.
This is also true for standard FisherRF, except for depth UE on LF.
The improvement is especially notable for rendering UE, which increases our method's UE performance to a level similar to its depth UE performance in the object-centric setting.
As expected, we observe universal improvements when training the meta-regression models on more than one holdout view.
As before, direction-dependent uncertainty feature maps perform slightly better for rendering UE.
However, this is only the case for the entire scene setting.
For all object-centric settings, our method achieves AUSE scores below \(0.1\) and correlations above \(0.8\) when trained on only one hold-out view, demonstrating robust UE performance with limited supervision.

For an analysis on how well PRIMU meta-regression models generalize to novel scenes, see Appendix~\ref{appen:UEoutOfSceneGeneralization}.
In short, we can achieve high UE performance in the object-centric setting when using regression models trained on multiple scenes and evaluated on novel ones.
Then, we also achieve high performance in depth UE and decent performance in rendering UE.

\begin{figure}
    \centering
    \begin{subfigure}[c]{0.47\textwidth}
        \includegraphics[width=0.47\linewidth]{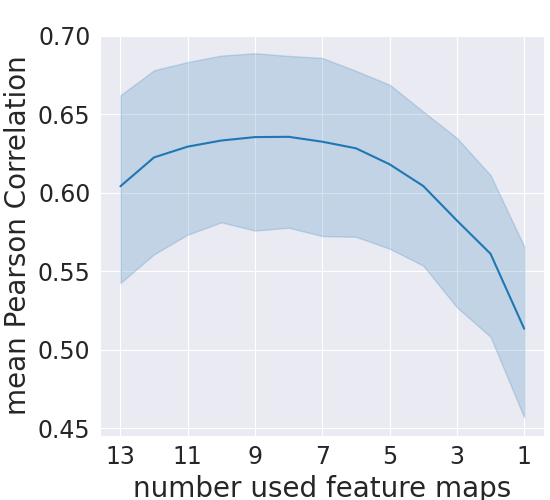}
        \includegraphics[width=0.51\linewidth]{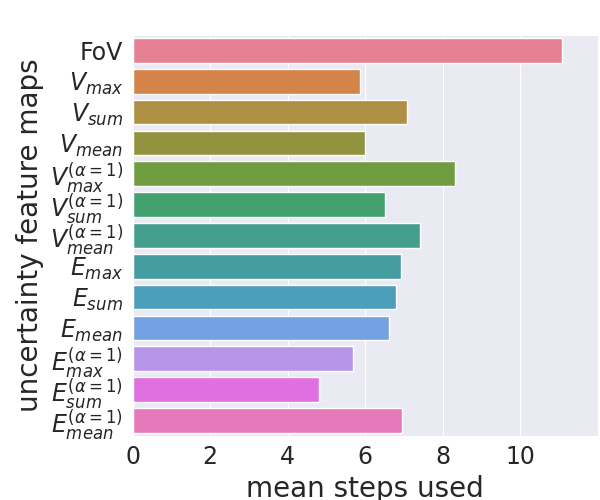}
        \subcaption{Entire scene evaluation}
    
        \includegraphics[width=0.47\linewidth]{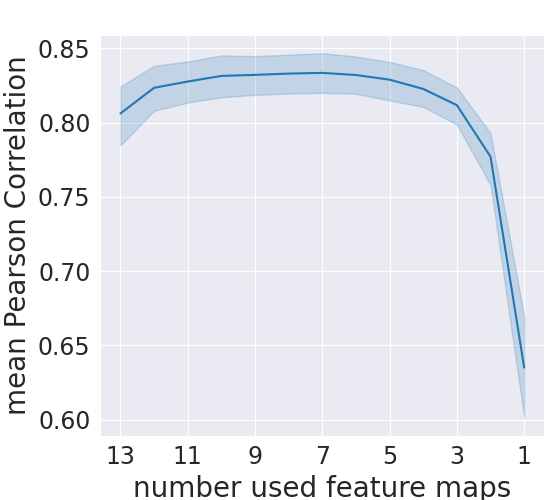}
        \includegraphics[width=0.51\linewidth]{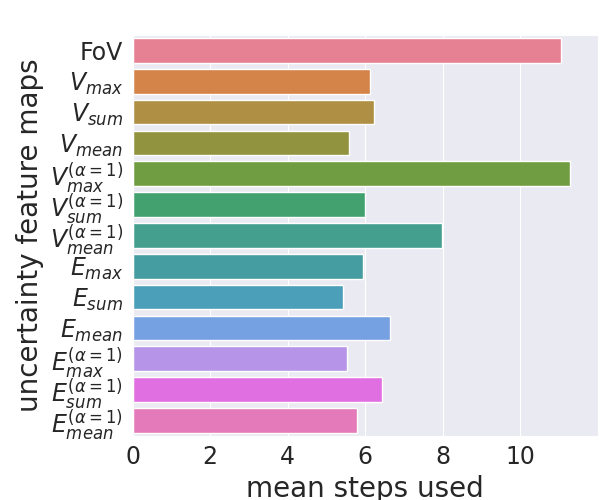}
        \subcaption{Object-centric evaluation}
    \end{subfigure}
    \caption{
    Step-wise backward regression results identifying the most informative uncertainty feature maps.
    Left: Pearson correlation between predicted and true depth error over the number of feature maps used.
    Right: average feature map retention times.
    }
    \label{fig:backRegRGBtrajectory}
\end{figure}

\paragraph{Uncertainty Feature Map Combination Study}\label{par:backReg}
We further examine which combinations of uncertainty feature maps contribute most to depth UE performance using step-wise backward regression.
This means we start training our regression models on all possible combinations of \(12\) direction-independent uncertainty feature maps, iteratively dropping those with the lowest Pearson correlation between the uncertainty map and depth error when left out.
This process is repeated for all scenes from LF, as well as for five scenes from TUM.
Regression models are trained on single hold-out views and evaluated on some of the remaining ones (2 for LF, 3 for TUM).

\Cref{fig:backRegRGBtrajectory} shows Pearson correlation trajectories, of uncertainties to depth error, and average feature retention times.
UE performance is higher and more stable in the object-centric case, with optimal performance reached when using about six maps versus nine for entire scenes.
The FoV counter and the coverage map with max aggregation (without opacity) consistently rank highest, highlighting their utility for UE and complementary roles: FoV reflects in how many training views a region of the splatting is included, while opacity-free coverage maps emphasize relevant but semi-transparent Gaussians.
In the entire scene setting, the FoV feature map is especially important, potentially because it allows one to distinguish regions that appear in many training views, such as the central object, from regions that appear in only a few training views, such as the background.
The importance of the coverage map without opacity and max aggregation, suggests that max aggregation leads to an effective estimate of generally high training-view coverage.
Error-based maps contribute less, likely due to being derived from rendering errors which are intuitively less informative for depth UE.

To assess whether reducing the number of uncertainty feature maps improves overall PRIMU performance, we selected some feature map combinations to be tested on all LF and TUM scenes following three strategies: a) the combination that was chosen most frequently across all stepwise backward regression runs, b) the combination with the highest overall score, and c) the set of individually most frequently selected uncertainty feature maps.
The number of feature maps is selected based on the average performance in the given setting.
In particular, six are selected for the object-centric case and nine for the entire scene setting.
Selected combinations (\cref{tab:fMapStudyResultsAll,tab:fMapStudyResultsObj}) maintain nearly full performance, indicating that there is some redundant information between the maps.
However, no reduced feature-map combination consistently improves PRIMU, with the full 13 maps typically yielding the best average performance.

\begin{table}
    \setlength{\tabcolsep}{2pt}
    \centering
    \scalebox{0.85}{
    \begin{tabular}{l|ll|ll|ll|ll}
\toprule
& \multicolumn{8}{c}{entire views} \\
& \multicolumn{4}{c|}{rendering error} & \multicolumn{4}{c}{depth error} \\
& \multicolumn{2}{c|}{LF} & \multicolumn{2}{c|}{TUM} & \multicolumn{2}{c|}{LF} & \multicolumn{2}{c}{TUM} \\
& AUSE & Pears. & AUSE & Pears. & AUSE & Pears. & AUSE & Pears. \\
\shortstack{no. of\\maps} & (\(\downarrow\)) & (\(\uparrow\)) & (\(\downarrow\)) & (\(\uparrow\)) & (\(\downarrow\)) & (\(\uparrow\)) & (\(\downarrow\)) & (\(\uparrow\)) \\
\midrule
13 & \textbf{0.244} & \textbf{0.412} & \underline{0.416} & \underline{0.217} & \textbf{0.112} & \textbf{0.734} & \textbf{0.213} & \textbf{0.536} \\
9a & \underline{0.245} & 0.407 & \textbf{0.413} & \textbf{0.222} & \underline{0.12} & \underline{0.721} & \underline{0.214} & \textbf{0.536} \\
9b & 0.248 & \underline{0.408} & 0.419 & 0.216 & 0.121 & 0.716 & 0.22 & 0.525 \\
9c & 0.249 & 0.402 & 0.423 & 0.211 & 0.121 & 0.708 & 0.216 & \underline{0.535} \\
\bottomrule
\end{tabular}

    }
    \caption{Feature map combination study for entire scenes.}
    \label{tab:fMapStudyResultsAll}
\end{table}

\begin{table}
    \setlength{\tabcolsep}{2pt}
    \centering
    \scalebox{0.85}{
    \begin{tabular}{l|ll|ll|ll|ll}
\toprule
& \multicolumn{8}{c}{object-centric} \\
& \multicolumn{4}{c|}{rendering error} & \multicolumn{4}{c}{depth error} \\
& \multicolumn{2}{c|}{LF} & \multicolumn{2}{c|}{TUM} & \multicolumn{2}{c|}{LF} & \multicolumn{2}{c}{TUM} \\
& AUSE & Pears. & AUSE & Pears. & AUSE & Pears. & AUSE & Pears. \\
\shortstack{no. of\\maps} & (\(\downarrow\)) & (\(\uparrow\)) & (\(\downarrow\)) & (\(\uparrow\)) & (\(\downarrow\)) & (\(\uparrow\)) & (\(\downarrow\)) & (\(\uparrow\)) \\
\midrule
13 & \textbf{0.072} & \textbf{0.822} & \textbf{0.101} & \textbf{0.756} & \underline{0.071} & \textbf{0.835} & \textbf{0.1} & \underline{0.745} \\
6a & 0.074 & 0.807 & 0.104 & 0.747 & \underline{0.071} & 0.822 & 0.105 & 0.736 \\
6b & 0.076 & 0.8 & \underline{0.103} & \underline{0.75} & 0.072 & 0.822 & 0.105 & 0.744 \\
6c & \underline{0.073} & \underline{0.811} & \underline{0.103} & \underline{0.75} & \textbf{0.069} & \underline{0.832} & \underline{0.101} & \textbf{0.746} \\
\bottomrule
\end{tabular}

    }
    \caption{Feature map combination study for object-centric scenes.}
    \label{tab:fMapStudyResultsObj}
\end{table}

\begin{table}
    \centering
    \setlength{\tabcolsep}{4pt}
    \scalebox{0.85}{
        \begin{tabular}{l|lll}
\toprule
 & PSNR & SSIM & LPIPS \\
method & (\(\uparrow\)) & (\(\uparrow\)) & (\(\downarrow\)) \\
\midrule
manifold & 19.620 & 0.592 & 0.376 \\
manifold$^\dagger$ & 20.102 & 0.610 & 0.351 \\
FisherRF & 20.392 & 0.586 & 0.363 \\
PRIMU-AVS (ours) & \textbf{20.721} & \textbf{0.625} & \textbf{0.34} \\
\bottomrule
\end{tabular}
    }
    \caption{
    AVS results on MipNeRF360. 
    PRIMU-AVS employs the direction-dependent coverage feature map with max aggregation and dropped opacity to directly guide view selection.
    }
    \label{tab:avsResults}
\end{table}

\paragraph{Experimental Setup for Active View Selection}
In GS, AVS aims to identify training views that yield the highest final reconstruction quality, while limiting the total number of training views.
The goal of our setup is to iteratively select the best \(20\) training views.
To ensure comparability with the FisherRF baseline, we adopt the AVS protocol proposed by Jiang et~al.~\cite{Jiang2023FisherRF}.

Following this setup, every eighth view is removed from the set of candidate training views and reserved for evaluation.
After selecting a fixed first training view, three additional views are selected by maximizing the Euclidean distance to the camera centers of the existing views. All methods start the training with these four views.
Additional, views are then selected by the AVS scheme during training, with each selection occurring after a number of iterations proportional to the number of views already selected.
This continues until \(20\) training views are reached.
Specifically, a new view is added after \(100\) times the number of selected training views (e.g., the fifth view after \(400\) iterations, the sixth after an additional \(500\), and so on).

Experiments are conducted on the challenging real-world MipNeRF360 dataset~\cite{barron2022mipnerf}. We use the standard GS training configuration from~\cite{kerbl20233dgaussiansplatting}, but increase the interval for raising the spherical harmonics degree during training from \(1000\) to \(5000\) iterations, following~\cite{Jiang2023FisherRF}, to mitigate overfitting caused by the limited number of views.

As baselines, we include the GS-based AVS methods by Jiang et al.~\cite{Jiang2023FisherRF} (FisherRF) and Lyu et al.~\cite{lyu2024manifold} (manifold). Since manifold uses a stochastic variant of GS, which could impact the reconstruction quality, we additionally provide results for a modified version denoted manifold$^*$. In manifold$^\dagger$, we use the views selected by manifold but in the vanilla GS~\cite{kerbl20233dgaussiansplatting} pipeline, which we also use for FisherRF and our approach.
Because AVS performance is evaluated through final rendering quality, comparisons to NeRF-based AVS methods are omitted, as NeRF and GS in general differ substantially in reconstruction quality.

In our AVS approach~(PRIMU-AVS), we omit the regression model since training it would require at least one additional view.
Instead, we use a single uncertainty feature map that is based on the direction-dependent coverage representation with \(\max\) aggregation and dropped opacity (\({V_{\max}^{(\alpha=1)}}^*\)).  
This selection scheme aims to maximize the total coverage of Gaussian primitives across the chosen training views.
The direction-dependent formulation allows the model to account for how well Gaussian primitives are observed from different viewing angles.
We employ the \(\max\) aggregation as it is conceptually simple and independent of the primitives’ spatial extent.
To ensure that transparent primitives are also considered when estimating coverage, we drop the opacity term.  
When selecting a new training view, we render the chosen coverage feature map for all candidate views and select the one with the lowest mean value across pixels, corresponding to the view with the lowest overall coverage
from the already selected training views.

\vspace{-0.1cm}
\paragraph{Active View Selection Results}
In \cref{tab:avsResults} we provide the final splatting quality, on the MipNeRF360 dataset, after the AVS training is completed.
As can be seen, PRIMU-AVS outperforms both FisherRF and manifold in the three standard GS quality metrics PSNR, SSIM and LIPIPS.
This demonstrates the effectiveness of our approach which uses uncertainty feature information based on training-view coverage in AVS.
We reliably achieve splatting quality that surpasses baseline methods which explicitly model uncertainty or estimate information gain.

\section{Conclusion}
\label{sec:conclusion}

We propose PRIMU, an effective post-hoc technique for uncertainty estimation in Gaussian splatting that achieves state-of-the-art performance for both depth and RGB rendering error prediction. PRIMU uses 13 handcrafted Gaussian primitive-based feature maps, capturing training-view coverage and projected rendering errors, and trains lightweight gradientboosting regressors on a single hold-out view. By leveraging PRIMU uncertainty feature maps directly, we achieve state-of-the-art results in active view selection.

\section*{Acknowledgment}
T.G.\ P.S., and M.R.\ acknowledge support by the state of North Rhine-Westphalia and the European Union within the EFRE/JTF project ``Just scan it 3D'', grant no.\ EFRE-20800529.
E.H.\ and M.R.\ acknowledge support through the junior research group project ``UnrEAL'' by the German Federal Ministry of Education and Research (BMBF), grant no.\ 01IS22069.

\FloatBarrier
{
    \small
    \bibliographystyle{ieeenat_fullname}
    \bibliography{ref}
}

\setcounter{page}{1}
\maketitlesupplementary
\appendix

This appendix provides additional analyses, implementation details, and qualitative results that complement the main paper.  
\cref{appen:fisherRFbaseline} describes our modification of the FisherRF baseline, introducing a regression-based variant used in our comparisons.  
\cref{appen:UEsceneSeparation} presents results of our scene-separation study, focusing specifically on uncertainty estimation for background regions.  
\cref{appen:UEoutOfSceneGeneralization} evaluates the out-of-scene generalization capabilities of the meta-regression models used in PRIMU.  
\cref{appen:VisualExamples} provides qualitative examples and visual comparisons of predicted uncertainty maps.
Finally, \cref{appen:UEspacialInteraction} examines the benefit of incorporating neighboring-pixel information for improving UE.

\section{Regression-Based FisherRF}
\label{appen:fisherRFbaseline}
This section outlines the regression-based FisherRF method used in the study of UE scene separation.
Original FisherRF \cite{Jiang2023FisherRF} is based on the Fisher information matrix of a GS w.r.t.\ the gradient of the rendering function.
The diagonal entries of this Fisher information matrix correspond to variances of the Gaussian primitive parameters and therefore are used to estimate their uncertainty.
The considered Gaussian primitive uncertainty of FisherRF only uses the sum of the color parameter variances including all spherical harmonic parameters, used to make the color direction-dependent. 
Using the GS rending (\cref{eq:GSrendering}), pixel-wise uncertainties are obtained.

We have made some modifications to make FisherRF more similar to our own method, so that it also uses a meta-regression model.
That is, we compute \(6\) uncertainty feature maps based on the Gaussian parameter variances like in FisherRF but for different Gaussian primitive parameters.
The \(6\) uncertainty feature maps are based on the grouping of parameters of Gaussian primitives in 1) mean, 2) scale, 3) rotation, 4) opacity factor,
5) spherical harmonics color parameters of degree zero and 6) the higher degree spherical harmonics color parameters.
This grouping of parameters originates from the original GS implementation \cite{kerbl20233dgaussiansplatting}.
For a Gaussian primitive the parameter groups scale and rotation correspond to its covariance matrix.
In this way all Gaussian primitive parameters are represented in one of the \(6\) uncertainty feature maps.
We use those \(6\) uncertainty feature maps as input for regression models in the same manner as for our uncertainty feature maps obtained by our method.
This allows us to compare the quality of the different sets of feature maps for UE by evaluating the pixel-wise predicted errors using our evaluation metrics.

\section{UE Scene Separation Study for Background}
\label{appen:UEsceneSeparation}

Here, we are extending our UE scene separation study.
We are also analyzing UE on the scene background alone.
Therefore, the meta-regression models for our method and the regression-based FisherRF are trained on the scene background.
In \cref{tab:backgroundStudyResults} we provide the numerical results.
For easier comparison, we also include the results for the entire scene again.
We observe that the effects on the scene background are analogous to those on the entire scene.
For instance, the best-performing configurations are identical.
This is to be expected, as the background usually dominates in terms of pixel counts in an image.
Typically, this is true when capturing an object in an environment, as in the scene datasets used here.
In general, the UE performance on the scene background is slightly worse than in the entire-scene setting.
Nevertheless, the overall performance of our method is good for the background, especially for depth UE.

\begin{table*}
    \setlength{\tabcolsep}{2pt}
    \centering
    \scalebox{0.85}{
    \begin{tabular}{l|ll|ll|ll|ll|ll|ll|ll|ll|ll}
\toprule
& \multicolumn{2}{r|}{predicted error type} & \multicolumn{8}{c|}{rendering error} & \multicolumn{8}{c}{depth error} \\
& \multicolumn{2}{r|}{evaluation on} & \multicolumn{4}{c|}{entire views} & \multicolumn{4}{c|}{background} & \multicolumn{4}{c|}{entire views} & \multicolumn{4}{c}{background} \\
 &  & dataset & \multicolumn{2}{c|}{LF} & \multicolumn{2}{c|}{TUM} & \multicolumn{2}{c|}{LF} & \multicolumn{2}{c|}{TUM} & \multicolumn{2}{c|}{LF} & \multicolumn{2}{c|}{TUM} & \multicolumn{2}{c|}{LF} & \multicolumn{2}{c}{TUM} \\
 &  & metric & AUSE & Pears. & AUSE & Pears.  & AUSE & Pears.  & AUSE & Pears.  & AUSE & Pears.  & AUSE & Pears.  & AUSE & Pears.  & AUSE & Pears.  \\
\multirow[t]{2}{*}{\rotatebox{90}{views}} & method & model & (\(\downarrow\)) & (\(\uparrow\)) & (\(\downarrow\)) & (\(\uparrow\)) & (\(\downarrow\)) & (\(\uparrow\)) & (\(\downarrow\)) & (\(\uparrow\)) & (\(\downarrow\)) & (\(\uparrow\)) & (\(\downarrow\)) & (\(\uparrow\)) & (\(\downarrow\)) & (\(\uparrow\)) & (\(\downarrow\)) & (\(\uparrow\)) \\
\midrule
 & FisherRF &  & 0.878 & -0.224 & 0.957 & -0.241 & 0.962 & -0.265 & 0.976 & -0.255 & 0.322 & 0.31 & 0.742 & -0.08 & 0.361 & 0.231 & 0.758 & -0.169 \\
\cline{1-19} \cline{2-19}
\multirow[t]{6}{*}{\rotatebox[origin=r]{90}{1 view}} & \multirow[t]{2}{*}{reg. FisherRF} & grad. & 0.282 & 0.318 & 0.435 & 0.181 & 0.277 & 0.33 & 0.44 & 0.183 & 0.133 & 0.683 & 0.328 & 0.375 & 0.163 & 0.627 & 0.328 & 0.357 \\
 &  & lin. & 0.332 & 0.182 & 0.404 & 0.143 & 0.311 & 0.198 & 0.41 & 0.143 & 0.212 & 0.452 & 0.363 & 0.305 & 0.265 & 0.363 & 0.357 & 0.283 \\
\cline{2-19}
 & \multirow[t]{2}{*}{PRIMU (ours)} & grad. & 0.244 & 0.412 & 0.416 & 0.217 & 0.251 & 0.41 & 0.435 & 0.199 & 0.112 & 0.734 & 0.213 & 0.536 & 0.136 & 0.697 & 0.235 & 0.497 \\
 &  & lin. & 0.286 & 0.309 & 0.353 & 0.29 & 0.287 & 0.335 & 0.369 & 0.279 & 0.202 & 0.545 & 0.188 & 0.571 & 0.254 & 0.468 & 0.213 & 0.529 \\
\cline{2-19}
 & \multirow[t]{2}{*}{PRIMU*(ours)} & grad. & 0.238 & 0.419 & 0.415 & 0.23 & 0.245 & 0.42 & 0.426 & 0.219 & 0.108 & 0.739 & 0.239 & 0.501 & 0.134 & 0.699 & 0.272 & 0.443 \\
 &  & lin. & 0.276 & 0.35 & 0.362 & 0.296 & 0.273 & 0.377 & 0.373 & 0.282 & 0.202 & 0.557 & 0.226 & 0.521 & 0.253 & 0.477 & 0.26 & 0.456 \\
\cline{1-19} \cline{2-19}
\multirow[t]{3}{*}{\rotatebox[origin=r]{90}{m-view}} & reg. FisherRF & grad. & 0.26 & 0.352 & 0.36 & 0.267 & 0.255 & 0.365 & 0.369 & 0.275 & 0.127 & 0.715 & 0.226 & 0.515 & 0.155 & 0.667 & 0.226 & 0.501 \\
\cline{2-19}
 & PRIMU (ours) & grad. & \underline{0.222} & \underline{0.446} & \underline{0.335} & \underline{0.302} & \underline{0.226} & \underline{0.45} & \underline{0.353} & \underline{0.293} & \underline{0.099} & \underline{0.777} & \textbf{0.132} & \textbf{0.653} & \underline{0.119} & \underline{0.745} & \textbf{0.155} & \textbf{0.611} \\
\cline{2-19}
 & PRIMU*(ours) & grad. & \textbf{0.218} & \textbf{0.458} & \textbf{0.329} & \textbf{0.321} & \textbf{0.223} & \textbf{0.461} & \textbf{0.338} & \textbf{0.312} & \textbf{0.097} & \textbf{0.788} & \underline{0.161} & \underline{0.605} & \textbf{0.117} & \textbf{0.753} & \underline{0.183} & \underline{0.57} \\
\cline{1-19} \cline{2-19}
\bottomrule
\end{tabular}
    }
    \caption{
    Scene separation UE results for scene background. “entire views” evaluates on the full GS scene; “background” limits to the scene background. PRIMU* uses direction-dependent coverage and error maps.
    }
    \label{tab:backgroundStudyResults}
\end{table*}

\section{UE Out-of-Scene Generalization}
\label{appen:UEoutOfSceneGeneralization}

We study how well the PRIMU UE meta-regression models generalize to new scenes.
If they do, it would suggest that the uncertainty information in our feature maps is at least partially scene-independent.
For this study, we use the setup from our UE scene separation study to also investigate generalization in object-centric and background settings.
For comparison, we also conduct generalization experiments for the regression-based FisherRF approach.
We train the meta-regression models using one view of one scene, multiple views of one scene, or multiple views of multiple scenes from the same dataset.
The results are provided in \cref{tab:generalization}, which also includes a row of standard FisherRF results for comparison.
To improve readability, we only report the Pearson correlation of the uncertainty maps to the true error.
The rows of the table correspond to the mean correlations of different meta-regression models trained using different sets of views from the training datasets (given by column "training").
The columns correspond to the means of evaluation over scenes in the evaluation datasets (given by row "dataset").
These means never include regression models that are trained and evaluated on the same scene.

Examining the results, we see that our method consistently outperforms the standard FisherRF method, even in settings where we train regression models using holdout views from different scenes.
The only noticeable exception is depth UE on the LF dataset in the entire-scene and background settings.
In these settings, both our method and the regression-based FisherRF method perform significantly worse than when training on holdout views of the same scene.
This is likely due to an inconsistent depth scale across the LF scenes.
When we compare to the TUM dataset in the depth UE, entire-scene, or background settings, we see a less pronounced drop in performance, and the depth in the TUM dataset is scaled in meters for all scenes.
Interestingly, we also see that, in some cases, linear regression trained on a single hold-out view performs best, indicating weaker nonlinear effects in these settings and suggesting a tendency of gradient boosting to overfit the training scenes.
We also notice that, in a few cases, models trained on TUM scenes perform better on LF than models trained on LF.
This may be due to the fact that TUM has more scenes and more diverse hold-out views within those scenes than LF, covering a wider range of scenarios.
Overall, we observe that meta-regression models can generalize to a certain extent to unseen scenes, and models trained on multiple scenes demonstrate the best generalization capability.

\begin{table*}
    \setlength{\tabcolsep}{4pt}
    \centering
    \scalebox{0.85}{
    \begin{tabular}{l|lll|l|l|l|l|l|l|l|l|l|l|l|l}
\toprule
 & \multicolumn{3}{r|}{predicted error type} & \multicolumn{6}{c|}{rendering error} & \multicolumn{6}{c}{depth error} \\
 & \multicolumn{3}{r|}{evaluation on} & \multicolumn{2}{c|}{entire views} & \multicolumn{2}{c|}{object-centric} & \multicolumn{2}{c|}{background} & \multicolumn{2}{c|}{entire views} & \multicolumn{2}{c|}{object-centric} & \multicolumn{2}{c}{background} \\
 &  &  & dataset & LF & TUM & LF & TUM & LF & TUM & LF & TUM & LF & TUM & LF & TUM \\
 &  &  & metric & Pears. & Pears. & Pears. & Pears. & Pears. & Pears. & Pears. & Pears. & Pears. & Pears. & Pears. & Pears. \\
\multirow[t]{2}{*}{\rotatebox{90}{views}} & method & model & training & (\(\uparrow\)) & (\(\uparrow\)) & (\(\uparrow\)) & (\(\uparrow\)) & (\(\uparrow\)) & (\(\uparrow\)) & (\(\uparrow\)) & (\(\uparrow\)) & (\(\uparrow\)) & (\(\uparrow\)) & (\(\uparrow\)) & (\(\uparrow\)) \\
\midrule
 & FisherRF &  &  & -0.224 & -0.241 & 0.129 & 0.458 & -0.265 & -0.255 & \underline{0.31} & -0.08 & 0.05 & 0.303 & \textbf{0.231} & -0.169 \\
\cline{1-16} \cline{2-16} \cline{3-16}
\multirow[t]{12}{*}{\rotatebox[origin=r]{90}{1 view}} & \multirow[t]{4}{*}{reg. FisherRF} & \multirow[t]{2}{*}{grad.} & LF & 0.153 & 0.163 & 0.328 & 0.42 & 0.168 & 0.157 & 0.24 & -0.008 & 0.284 & 0.312 & \underline{0.176} & -0.095 \\
 &  &  & TUM & 0.104 & 0.13 & 0.273 & 0.419 & 0.126 & 0.137 & 0.139 & 0.273 & 0.245 & 0.278 & 0.039 & 0.245 \\
\cline{3-16}
 &  & \multirow[t]{2}{*}{lin.} & LF & 0.115 & 0.078 & 0.184 & 0.386 & 0.117 & 0.065 & 0.19 & 0.123 & 0.259 & 0.253 & 0.107 & 0.086 \\
 &  &  & TUM & 0.076 & 0.099 & 0.087 & 0.329 & 0.081 & 0.097 & 0.15 & 0.203 & 0.089 & 0.184 & 0.066 & 0.175 \\
\cline{2-16} \cline{3-16}
 & \multirow[t]{4}{*}{PRIMU (ours)} & \multirow[t]{2}{*}{grad.} & LF & 0.192 & 0.168 & 0.509 & 0.481 & 0.199 & 0.124 & 0.121 & 0.153 & 0.632 & 0.453 & 0.017 & 0.059 \\
 &  &  & TUM & 0.122 & 0.177 & 0.398 & 0.484 & 0.141 & 0.158 & 0.187 & 0.404 & 0.514 & 0.506 & 0.049 & 0.335 \\
\cline{3-16}
 &  & \multirow[t]{2}{*}{lin.} & LF & 0.215 & 0.197 & 0.206 & 0.013 & 0.225 & 0.198 & 0.25 & 0.097 & 0.444 & 0.128 & 0.105 & 0.063 \\
 &  &  & TUM & 0.186 & 0.239 & 0.165 & 0.296 & 0.204 & 0.227 & \textbf{0.312} & 0.468 & 0.265 & 0.257 & 0.076 & 0.392 \\
\cline{2-16} \cline{3-16}
 & \multirow[t]{4}{*}{PRIMU*(ours)} & \multirow[t]{2}{*}{grad.} & LF & 0.2 & 0.167 & 0.548 & 0.506 & 0.196 & 0.095 & 0.159 & 0.104 & 0.631 & 0.452 & 0.031 & 0.016 \\
 &  &  & TUM & 0.155 & 0.203 & 0.424 & 0.5 & 0.166 & 0.18 & 0.183 & 0.378 & 0.513 & 0.5 & 0.048 & 0.3 \\
\cline{3-16}
 &  & \multirow[t]{2}{*}{lin.} & LF & \textbf{0.288} & 0.202 & 0.179 & 0.1 & \textbf{0.297} & 0.18 & 0.242 & 0.076 & 0.384 & 0.17 & 0.126 & 0.059 \\
 &  &  & TUM & 0.248 & 0.259 & 0.067 & 0.262 & 0.247 & 0.247 & 0.281 & 0.413 & 0.123 & 0.166 & 0.078 & 0.331 \\
\cline{1-16} \cline{2-16} \cline{3-16}
\multirow[t]{6}{*}{\rotatebox[origin=r]{90}{m-view}} & \multirow[t]{2}{*}{reg. FisherRF} & \multirow[t]{2}{*}{grad.} & LF & 0.15 & 0.16 & 0.343 & 0.448 & 0.17 & 0.164 & 0.243 & -0.01 & 0.284 & 0.315 & 0.159 & -0.106 \\
 &  &  & TUM & 0.145 & 0.186 & 0.308 & 0.445 & 0.175 & 0.198 & 0.193 & 0.349 & 0.31 & 0.304 & 0.072 & 0.305 \\
\cline{2-16} \cline{3-16}
 & \multirow[t]{2}{*}{PRIMU (ours)} & \multirow[t]{2}{*}{grad.} & LF & 0.194 & 0.195 & 0.491 & 0.513 & 0.2 & 0.128 & 0.13 & 0.148 & 0.657 & 0.464 & 0.017 & 0.084 \\
 &  &  & TUM & 0.13 & 0.226 & 0.444 & 0.5 & 0.155 & 0.208 & 0.203 & 0.477 & 0.584 & 0.577 & 0.057 & 0.412 \\
\cline{2-16} \cline{3-16}
 & \multirow[t]{2}{*}{PRIMU*(ours)} & \multirow[t]{2}{*}{grad.} & LF & 0.206 & 0.176 & 0.551 & 0.497 & 0.203 & 0.074 & 0.153 & 0.12 & 0.644 & 0.448 & 0.038 & -0.014 \\
 &  &  & TUM & 0.199 & 0.257 & 0.497 & 0.524 & 0.206 & 0.229 & 0.247 & 0.45 & 0.6 & 0.585 & 0.079 & 0.377 \\
\cline{1-16} \cline{2-16} \cline{3-16}
\multirow[t]{6}{*}{\rotatebox[origin=r]{90}{m-scene}} & \multirow[t]{2}{*}{reg. FisherRF} & \multirow[t]{2}{*}{grad.} & LF & 0.248 & 0.217 & 0.532 & 0.503 & 0.248 & 0.201 & 0.083 & -0.046 & 0.261 & 0.356 & 0.062 & -0.093 \\
 &  &  & TUM & 0.213 & 0.261 & 0.434 & 0.502 & 0.244 & 0.268 & 0.201 & 0.456 & 0.419 & 0.418 & 0.058 & 0.423 \\
\cline{2-16} \cline{3-16}
 & \multirow[t]{2}{*}{PRIMU (ours)} & \multirow[t]{2}{*}{grad.} & LF & 0.264 & 0.202 & 0.575 & 0.454 & 0.268 & 0.135 & 0.039 & 0.095 & \underline{0.687} & 0.455 & 0.009 & 0.036 \\
 &  &  & TUM & 0.165 & \underline{0.305} & 0.544 & \underline{0.559} & 0.184 & \underline{0.291} & 0.217 & \textbf{0.561} & 0.642 & \underline{0.628} & 0.1 & \textbf{0.488} \\
\cline{2-16} \cline{3-16}
 & \multirow[t]{2}{*}{PRIMU*(ours)} & \multirow[t]{2}{*}{grad.} & LF & \underline{0.266} & 0.112 & \textbf{0.611} & 0.517 & 0.266 & 0.045 & -0.03 & -0.004 & 0.663 & 0.463 & -0.051 & -0.002 \\
 &  &  & TUM & 0.256 & \textbf{0.333} & \underline{0.583} & \textbf{0.572} & \underline{0.269} & \textbf{0.315} & 0.274 & \underline{0.545} & \textbf{0.688} & \textbf{0.645} & 0.133 & \underline{0.474} \\
\cline{1-16} \cline{2-16} \cline{3-16}
\bottomrule
\end{tabular}

    }
    \caption{
    Numerical results of UE out-of-scene generalization study.
    Pearson Correlation of predicted to true depth/rendering error for entire/object-centric/background scene evaluation.
    All rows except the first are the averages when training the regressor on the dataset in the column 'training'.
    PRIMU* uses direction-dependent coverage and error maps.
    }
    \label{tab:generalization}
\end{table*}

\section{Visual UE Examples}
\label{appen:VisualExamples}
\begin{figure*}
    \centering
    \footnotesize
    \setlength{\tabcolsep}{0pt}
    \begin{tabular}{>{\centering\arraybackslash} m{0.142\linewidth} >{\centering\arraybackslash} m{0.142\linewidth} >{\centering\arraybackslash} m{0.142\linewidth} >{\centering\arraybackslash} m{0.142\linewidth} >{\centering\arraybackslash} m{0.142\linewidth} >{\centering\arraybackslash} m{0.142\linewidth} >{\centering\arraybackslash} m{0.142\linewidth}}
         Ground Truth & Rendering & \(\ell_1\) Error Map & PRIMU* & manifold & var3DGS & FisherRF
    \end{tabular}
    \includegraphics[width=\linewidth,trim={0 0.5cm 0 0.5cm},clip]{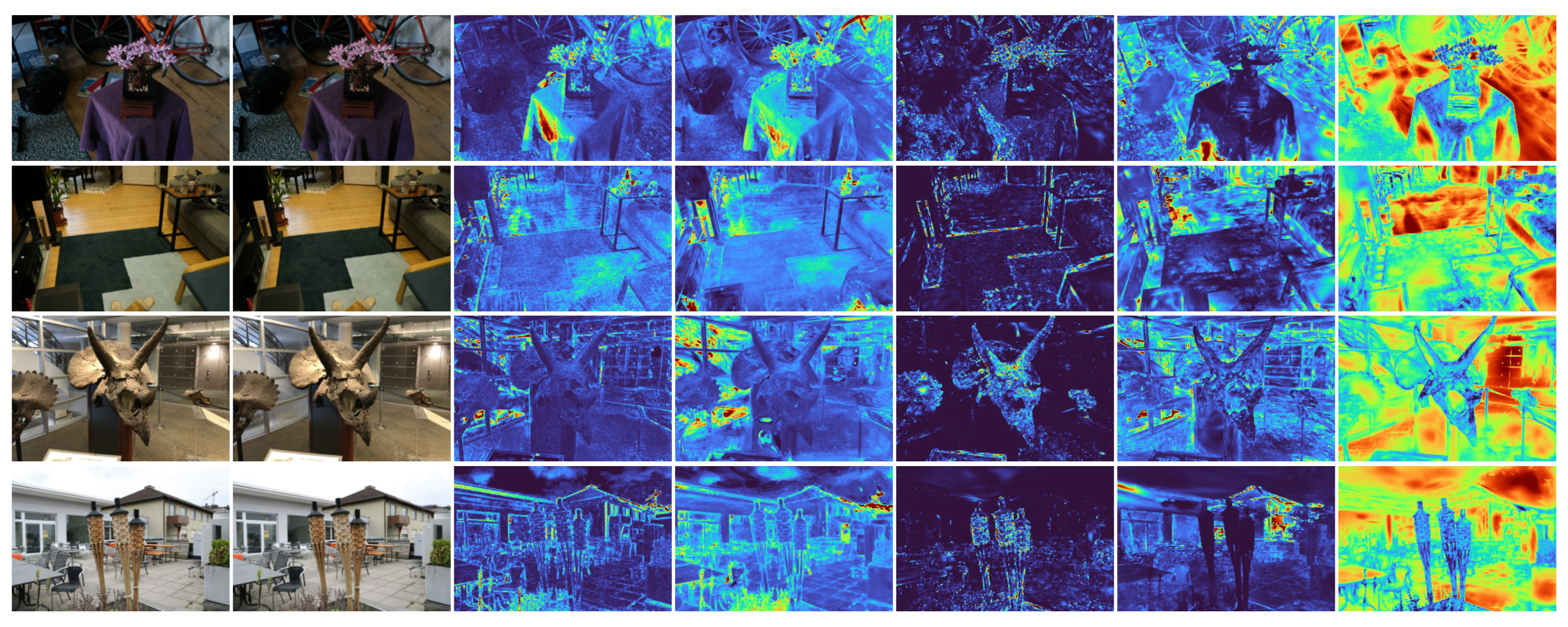}
    \caption{Qualitative comparison of uncertainty feature maps for rendering error UE of FisherRF~\cite{Jiang2023FisherRF}, manifold~\cite{lyu2024manifold}, var3DGS~\cite{li2024variational} and direction-dependent PRIMU* (ours), 
    to the \(\ell_1\) rendering error.
    Scenes from top to bottom: MipNeRF360 bonsai and room, LLFF horns and LF torch.}
    \label{fig:qualtiative_comparision2}
\end{figure*}

\begin{figure*}
    \centering
    \footnotesize
    \setlength{\tabcolsep}{0pt}
    \begin{tabular}{>{\centering\arraybackslash} m{0.142\linewidth} >{\centering\arraybackslash} m{0.142\linewidth} >{\centering\arraybackslash} m{0.142\linewidth} >{\centering\arraybackslash} m{0.142\linewidth} >{\centering\arraybackslash} m{0.142\linewidth} >{\centering\arraybackslash} m{0.142\linewidth} >{\centering\arraybackslash} m{0.142\linewidth}}
         Ground Truth & Rendering & \(\ell_1\) Error Map & PRIMU* & manifold & BayesRays & FisherRF
    \end{tabular}
    \includegraphics[width=\linewidth,trim={0 0.5cm 0 0.5cm},clip]{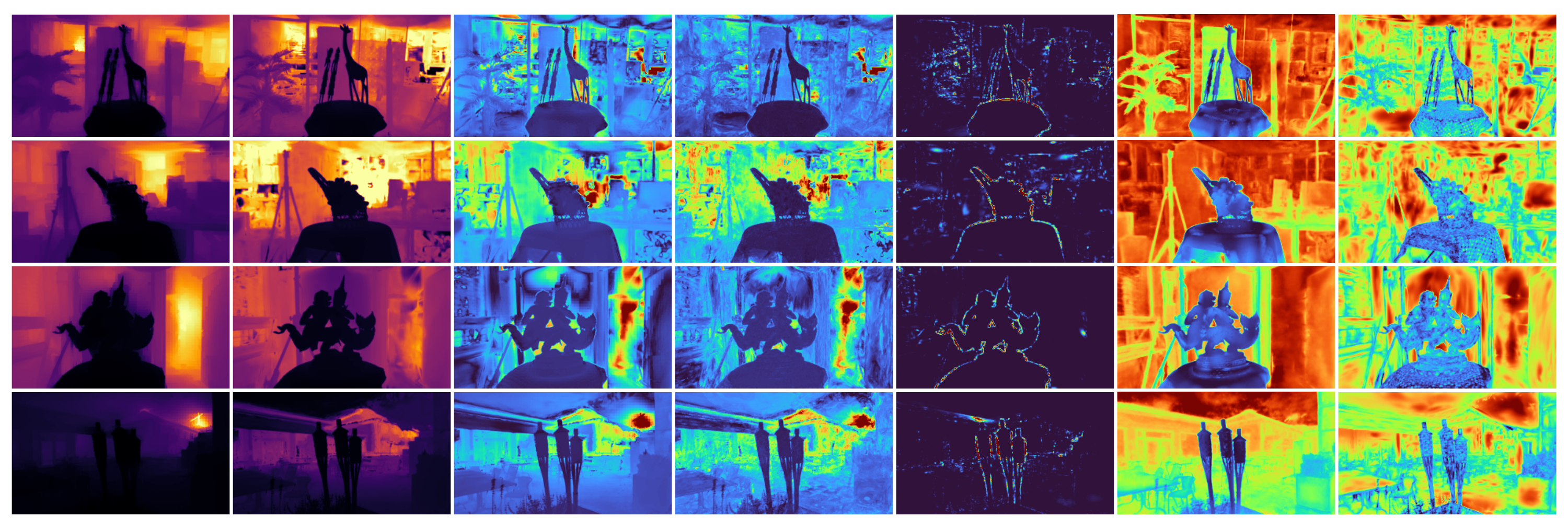}
    \caption{Qualitative comparison of uncertainty feature maps for depth UE of FisherRF~\cite{Jiang2023FisherRF}, manifold~\cite{lyu2024manifold}, BayesRays~\cite{goli2024bayesrays} and direction-dependent PRIMU* (ours), 
    to the \(\ell_1\) rendering error.
    Scenes from top to bottom: LF africa, basket, statue and torch.}
    \label{fig:qualtiative_comparision_depth}
\end{figure*}

\begin{figure*}
    \centering
    \footnotesize
    \begin{tabular}{cc|cc}
        \multicolumn{2}{c|}{UE for rendering error} &
        \multicolumn{2}{c}{UE for depth error} \\
        \midrule
        rendering & rendering error &
        rendered depth & depth error \\
        \includegraphics[width=0.22\linewidth]{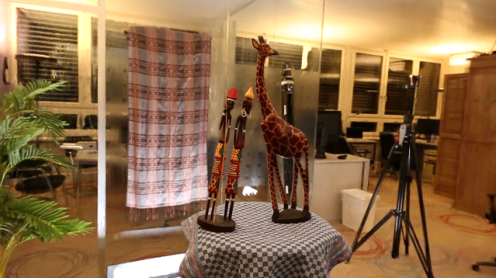} &
        \includegraphics[width=0.22\linewidth]{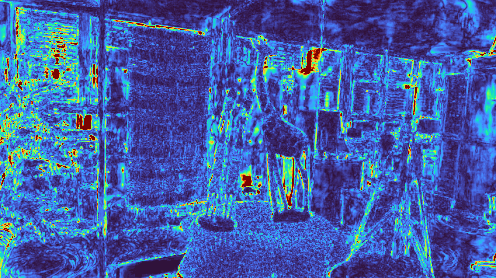} &
        \includegraphics[width=0.22\linewidth]{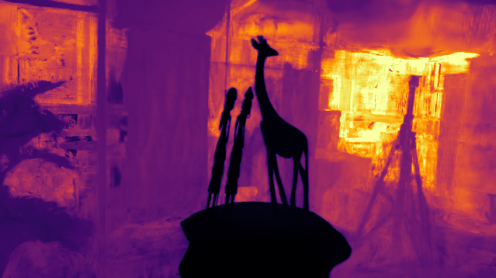} &
        \includegraphics[width=0.22\linewidth]{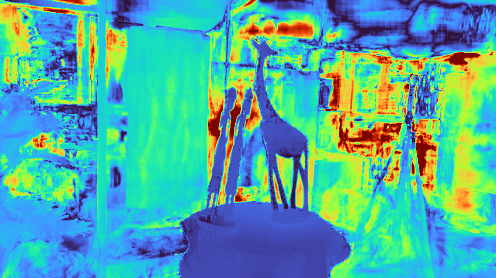} \\
        \bottomrule
    \end{tabular}
    \begin{tabular}{c|c|c|c}
        \toprule
        direction-independent & direction-dependent &
        direction-independent & direction-dependent \\
        \midrule
        \multicolumn{2}{c|}{gradient boosting trained on multiple views} &
        \multicolumn{2}{c}{gradient boosting trained on multiple views} \\
        \includegraphics[width=0.22\linewidth]{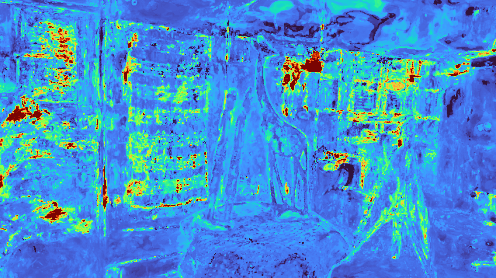} &
        \includegraphics[width=0.22\linewidth]{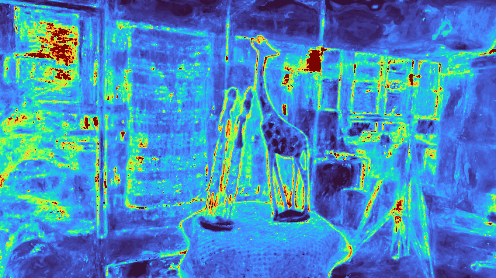} &
        \includegraphics[width=0.22\linewidth]{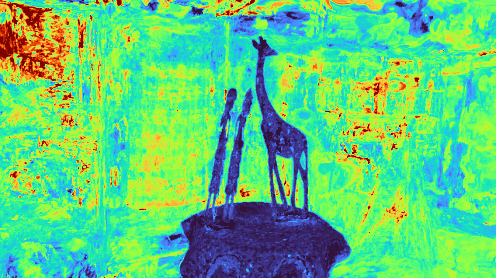} &
        \includegraphics[width=0.22\linewidth]{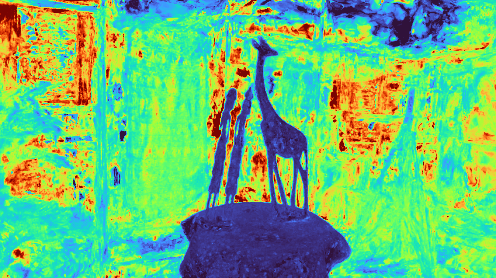} \\
    
        \midrule
        \multicolumn{2}{c|}{gradient boosting trained on one view} &
        \multicolumn{2}{c}{gradient boosting trained on one view} \\
        \includegraphics[width=0.22\linewidth]{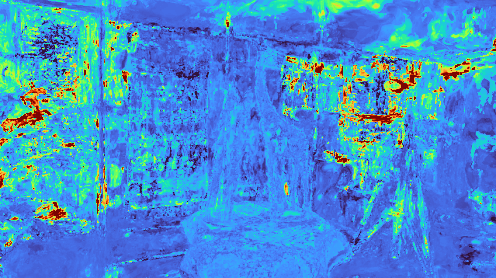} &
        \includegraphics[width=0.22\linewidth]{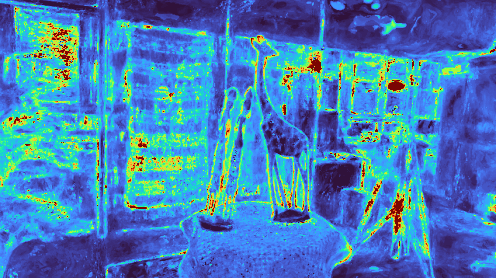} &
        \includegraphics[width=0.22\linewidth]{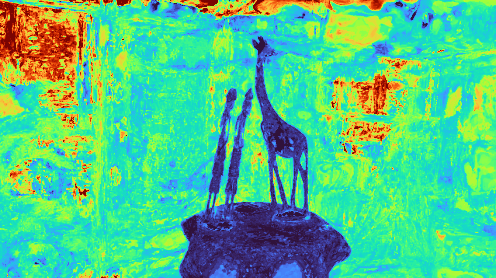} &
        \includegraphics[width=0.22\linewidth]{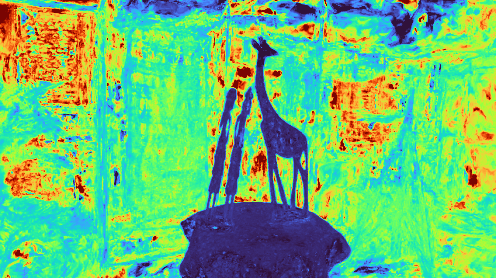} \\
        
        \midrule
        \multicolumn{2}{c|}{linear regression trained on one view} &
        \multicolumn{2}{c}{linear regression trained on one view} \\
         \includegraphics[width=0.22\linewidth]{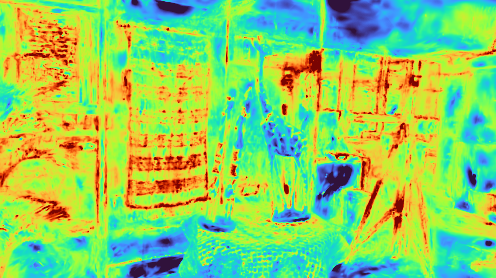} &
         \includegraphics[width=0.22\linewidth]{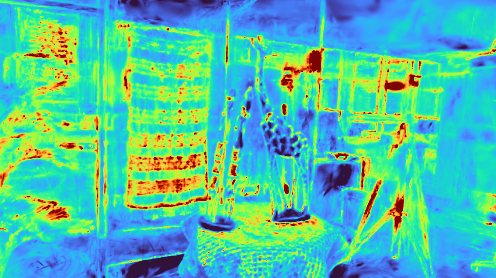} &
        \includegraphics[width=0.22\linewidth]{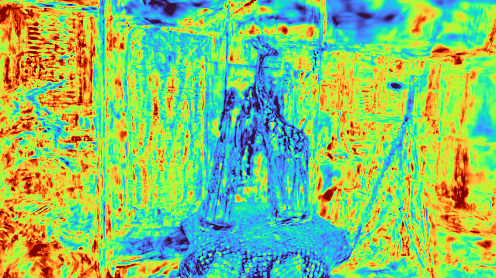} &
        \includegraphics[width=0.22\linewidth]{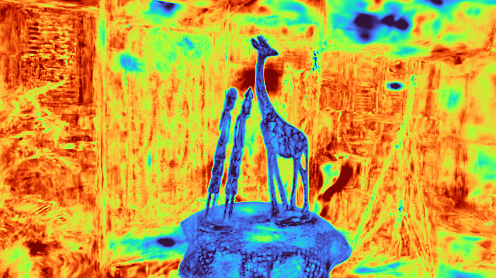}

    \end{tabular}
    \caption{
    Qualitative example of the UE of our method for the scene africa of the LF dataset.
    Top row color and depth renderings of view and their error maps.
    Below predicted error from different regression models trained on direction-independent or direction-dependent (plus FoV counter) uncertainty feature maps.
    }
    \label{fig:qualitativeExampleAfrica}
\end{figure*}

\begin{figure*}
    \centering
    \footnotesize
    \begin{tabular}{cc|cc}
        \multicolumn{2}{c}{UE for rendering error} &
        \multicolumn{2}{c|}{UE for depth error} \\
        \midrule
        rendering & rendering error &
        rendered depth & depth error \\
        \includegraphics[width=0.22\linewidth]{images/01_intro/fr3_teddy/fr3_teddy_9_all_rgb_render.png} &
        \includegraphics[width=0.22\linewidth]{images/01_intro/fr3_teddy/fr3_teddy_9_all_rgb_error.png} &
        \includegraphics[width=0.22\linewidth]{images/01_intro/fr3_teddy/fr3_teddy_9_all_depth_render.png} &
        \includegraphics[width=0.22\linewidth]{images/01_intro/fr3_teddy/fr3_teddy_9_all_depth_error.png} \\
        \bottomrule
    \end{tabular}
    \begin{tabular}{c|c|c|c}
        \toprule
        direction-independent & direction-dependent &
        direction-independent & direction-dependent \\
        \midrule
        \multicolumn{2}{c|}{gradient boosting trained on multiple views} &
        \multicolumn{2}{c}{gradient boosting trained on multiple views} \\
        \includegraphics[width=0.22\linewidth]{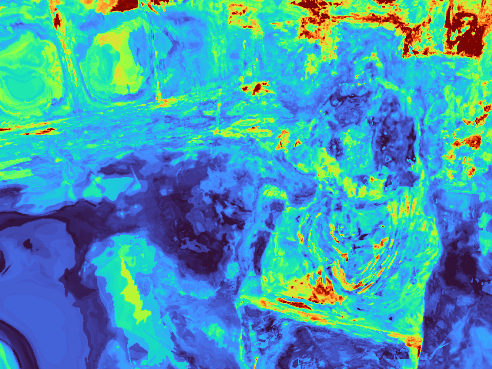} &
        \includegraphics[width=0.22\linewidth]{images/01_intro/fr3_teddy/run_regression_on_multiple_frames/fr3_teddy_9_all_rgb_pred_error.png} &
        \includegraphics[width=0.22\linewidth]{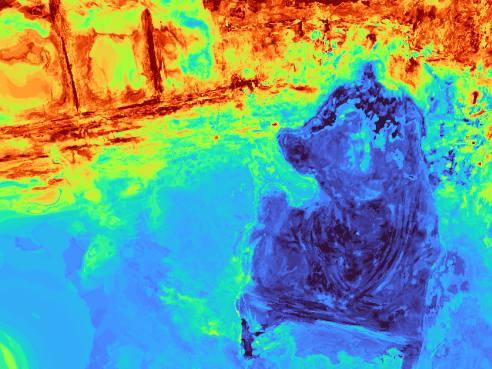} &
        \includegraphics[width=0.22\linewidth]{images/01_intro/fr3_teddy/run_regression_on_multiple_frames/fr3_teddy_9_all_depth_pred_error.png} \\
        
        \midrule
        \multicolumn{2}{c|}{gradient boosting trained on one view} &
        \multicolumn{2}{c}{gradient boosting trained on one view} \\
        \includegraphics[width=0.22\linewidth]{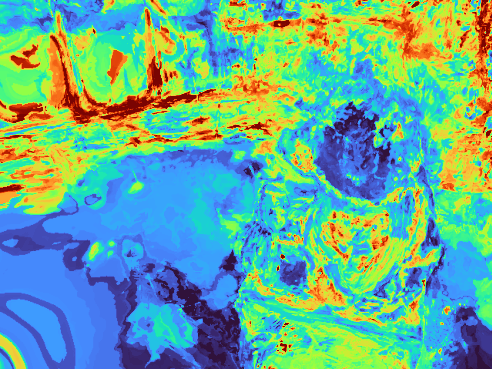} &
        \includegraphics[width=0.22\linewidth]{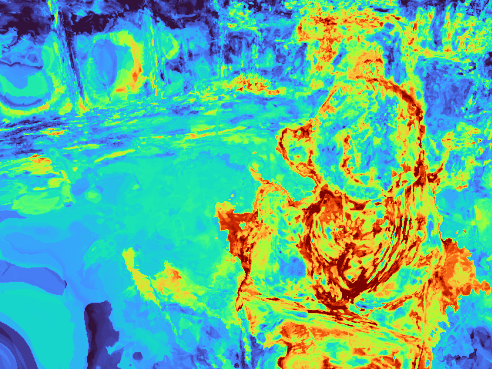} &
        \includegraphics[width=0.22\linewidth]{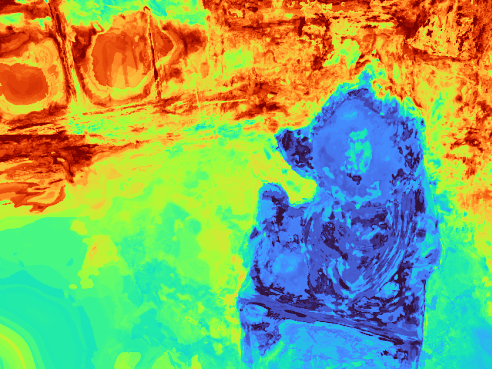} &
        \includegraphics[width=0.22\linewidth]{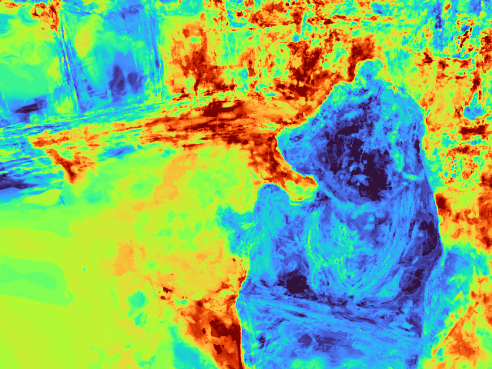} \\
        
        \midrule
        \multicolumn{2}{c|}{linear regression trained on one view} &
        \multicolumn{2}{c}{linear regression trained on one view} \\
        \includegraphics[width=0.22\linewidth]{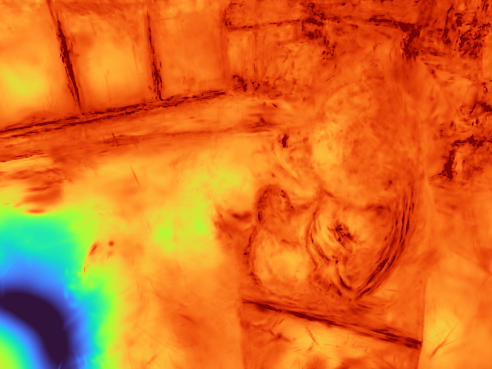} &
        \includegraphics[width=0.22\linewidth]{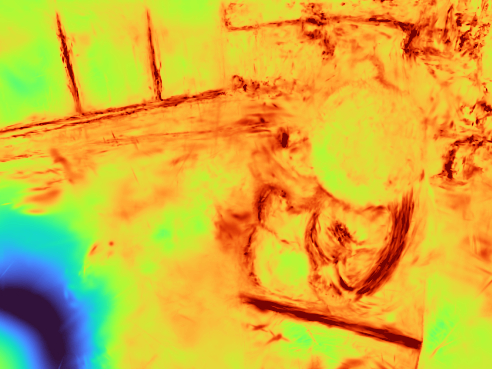} &
        \includegraphics[width=0.22\linewidth]{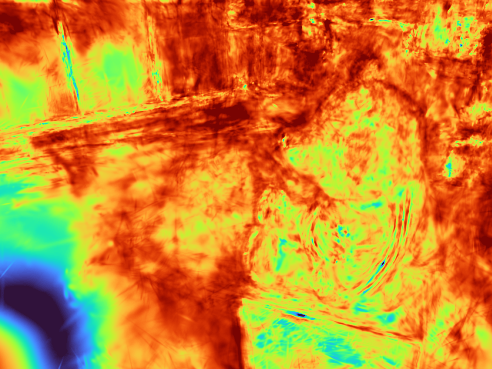} &
        \includegraphics[width=0.22\linewidth]{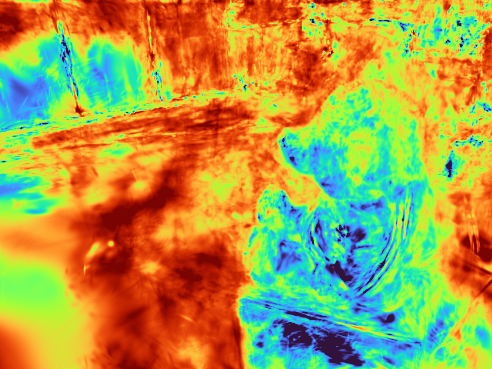}

    \end{tabular}
    \caption{
    Qualitative example of the UE of our method for the scene teddy2 of the TUM dataset.
    Top row color and depth renderings of view and their error maps.
    Below predicted error from different regression models trained on direction-independent or direction-dependent (plus FoV counter) uncertainty feature maps.
    }
    \label{fig:qualitativeExampleTeddy2}
\end{figure*}

Here, we provide some additional qualitative examples and comparisons of uncertainty maps.
In \cref{fig:qualtiative_comparision2}, we show more qualitative comparisons of the uncertainty maps of our method against the baselines for rendering error UE, and in \cref{fig:qualtiative_comparision_depth}, for depth UE.
\cref{fig:qualitativeExampleAfrica,fig:qualitativeExampleTeddy2} show a visual comparison of the uncertainty maps of our method for different choices of regression model, number of training views, and direction-dependent or independent uncertainty feature maps.
As we can see, the uncertainty maps produced by linear regression models are less nuanced than those produced by gradient boosting models.
This is likely due to the limited capacity of linear regression models compared to gradient boosting models.
Using direction-dependent uncertainty feature maps makes the predicted uncertainty maps more visually similar to the actual error maps.
For the Teddy2 scene from the TUM dataset (see \cref{fig:qualitativeExampleTeddy2}), we observe that training the regression model with multiple hold-out views also notably improves the visual quality of the uncertainty maps.
However, this is not the case for the Africa scene from LF (see \cref{fig:qualitativeExampleAfrica}), where the uncertainty maps produced by training on one hold-out view are already very close to the actual error maps.

\section{The Role of Spatial Interaction}
\label{appen:UEspacialInteraction}
\begin{figure}
    \centering
    \includegraphics[width=0.49\linewidth]{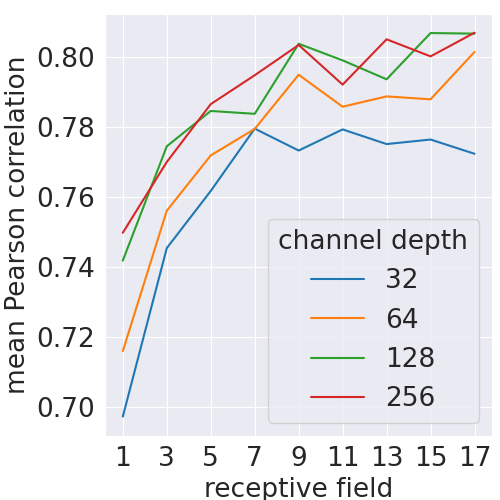}
    \includegraphics[width=0.49\linewidth]{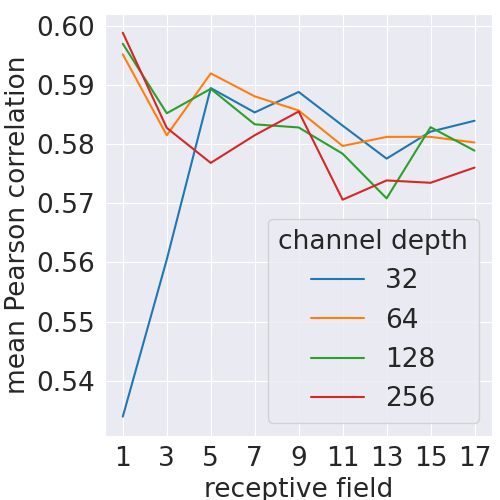}
    \caption{Average in-scene Pearson correlations of predicted and true rendering (left) and depth errors (right) for CNN models with varying receptive field sizes and numbers of uncertainty features.}
    \label{fig:cnn-results}
\end{figure}

So far, we performed pixel-wise regression on uncertainty feature maps. Here, we study whether incorporating uncertainty features from neighboring pixels improves the estimation of rendering and depth uncertainty. We replace the pixel-wise gradient boosting regressor with lightweight fully convolutional neural networks (CNNs) and evaluate their UE on the LF dataset. We train 72 CNNs to predict rendering and depth errors on foreground objects, using all 13 direction-independent primitive representations from the first three hold-out views of a scene and testing on the fourth view. Each CNN has a depth of 4 layers, with kernel sizes in 
\begin{equation}
    \{(k_1,\dots,k_4)| \, k_i \in \{1,3,5\}, \, k_j+2 \geq k_i \geq k_j \, \text{ for } \; i < j\}
\end{equation}
and channel depth in $\{(h,h,h,h)| h \in \{32,64,128,256\}\}$.  Notably, CNNs with all kernel sizes set to 1 reduce to pixel-wise fully connected baselines using only $1 \times 1$ convolutions. As batch size we found $1$ most useful and trained all CNNs for $2000$ iterations with a learning rate scheduled to decrease from $10^{-3}$ to $10^{-5}$, along with a weight decay of $10^{-5}$. UE performances are provided in terms of Pearson correlation between the predicted and true errors, displayed in \cref{fig:cnn-results}.

Overall, the CNNs did not outperform gradient boosting, which, however, is not the aim of this study. Interestingly, CNNs with larger receptive fields consistently outperform the fully-connected baselines in rendering error estimation, demonstrating the usefulness of neighborhood information for this task. For depth error estimation, increased receptive field size improves performance only for CNNs with 32 channels, which may be attributed to increased model capacity instead of neighborhood information itself.

\end{document}